# External anchors reveal target-population effects hidden by published clinical-trial evidence

**One-sentence summary:** An external distribution of untreated outcomes lets a meta-analysis estimate the effect in a declared trial-eligible patient population rather than the average over published trials, and supplies the information needed to correct publication selection—a strict holdout recovers the aggregate all-trials benchmark within uncertainty, without using the hidden trials in fitting or anchoring.

**Kasumi Dan**[1], **Ryoya Nakano**[1], **Takahiro Hoshino**[2,3*]

[1]Graduate School of Economics, Keio University, Tokyo, Japan. [2]Faculty of Economics, Keio University, Tokyo, Japan. [3]RIKEN Center for Advanced Intelligence Project, Tokyo, Japan.

*Corresponding author. Email: hoshino@econ.keio.ac.jp

---

**Abstract.** Meta-analyses guide clinical and policy decisions, but they estimate the effect among published trials, not the effect in the population a decision concerns. We introduce *reference-anchored meta-analysis*, which uses an external untreated-outcome distribution for two purposes: to define the target population, and, through each trial's control-arm mean as an externally anchored instrument, to correct publication selection. In an antidepressant literature with recovered unpublished trials, a strict holdout—removing the hidden trials from both fitting and anchoring—recovers a held-out all-trials benchmark. Applied to insomnia total sleep time and the placebo-controlled HbA1c literature in type-2 diabetes, it returns target-population effects materially smaller than the published averages, while a first-stage diagnostic flags, in advance, where the correction should not be attempted. Evidence synthesis becomes a target-explicit estimate with auditable selection assumptions.

## Introduction

Clinicians and policymakers read a meta-analysis as *the effect of a treatment*. What a meta-analysis mechanically returns, however, is an average over the studies that happened to be published—and that average can diverge from the quantity a decision requires for two distinct reasons. The first is *publication selection*: published trials are not a random sample of completed ones, and the tendency to publish on the direction or significance of a result is documented across nearly every empirical field, inflating pooled effects (*1*–*4*). The second, less often named, is *target mismatch*: even a selection-free synthesis estimates the effect at the particular mix of control conditions, severities, and populations that the assembled trials embody, which need not match the population a guideline or reimbursement decision applies to. Existing corrections address only the first problem—and, lacking any external target distribution, cannot separate selection from target mismatch—while under selection that depends on the unobserved effect size, the regime in which most methods become inconsistent, they address even that incompletely (*5*–*11*).

Here we show that a single externally specified object separates and addresses both problems under explicit anchor and exclusion assumptions—the anchor audited against external references and registries, the exclusion probed by sensitivity analysis: a *reference distribution* for the untreated, control-group outcome in the population that trials are meant to serve. Such distributions are available for many objective, normed, or criterion-referenced endpoints—sleep duration among adults with insomnia, HbA1c among adults with type-2 diabetes, and some clinician-rated symptom scales in well-characterized samples—from surveys, registries, or test norms that owe nothing to the published trial record.

The reference does double duty (Fig. 1). As the *target*, it fixes the population at which the effect is reported, converting the estimand from "the effect among published trials" into "the effect in the patient population." As an *anchor*, it identifies publication selection. The device is each trial's control-arm mean, which we use as a *shadow instrument*: a variable observed in published trials and anchored externally in the target population, associated with the reported effect—more severe baseline cases tend to yield larger standardized gains—yet, given that reported effect and standard design covariates, plausibly excluded from direct influence on whether a trial is published. The obstacle that has blocked such instruments before is that they go unobserved for precisely the trials that go unpublished (*12*, *13*); pinning the reference externally removes it, because the instrument's distribution is supplied from outside the selected sample. The published record alone cannot reveal *absolute* publication rates; those we calibrate separately from trial registries, against the broader denominator of a *retrievable result* (publication, registry posting, or regulatory filing) rather than journal publication alone. The selection the anchor identifies is *publication* selection insofar as the external control-level distribution represents the completed-trial-generating distribution before publication; where it does not, the method estimates selection of the published evidence relative to the declared target population, with the target shift reported separately.

Two commitments shape what follows. First, because the answer in a publication-bias problem is normally hidden, we test the method where the answer is observable: a strict holdout against a rare literature with recovered unpublished trials. Second, the method is built to *report its own inapplicability*—a first-stage diagnostic, computed before any correction, flags the literatures where the data cannot support it. The contribution is identification: an external reference, independent of the published record, supplies identifying content that published-only methods do not have. We address publication selection *across* trials, not selective outcome reporting *within* them, which is a distinct problem.

# Results

## An external untreated-outcome distribution does double duty

For each trial $k$ we observe a standardized treatment effect $Y_k$, its standard error $SE_k$, the control-arm outcome mean $C_k$ and arm size, and design covariates $W_k$. The reported effect varies with baseline severity, $Y_k = \alpha + \beta_C C_k + \beta_W W_k + u_k$, so the slope $\beta_C$ is the channel that makes the control mean informative about the effect. An external source supplies the mean and variance of the untreated outcome in the *trial-eligible* population, which a central-limit bridge maps to the study-level distribution of each observed control mean (Materials and Methods).

As an *anchor*, the deviation of published control means from the reference identifies selection: published trials whose control arms are systematically more (or less) severe than the trial-eligible population, in a way linked to the reported effect, reveal the selection that produced them. As a *target*, the reference fixes the control level at which the pooled effect is reported, defining the estimand

$$\Delta^* = \int \mathbb{E}_\theta[\,Y \mid C = c,\ U = u\,]\, m^*(c \mid u)\, dP^*(u)\, dc$$

the bias-corrected effect at the reference control level $m^*$, averaged over the trial-eligible covariate distribution. $\Delta^*$ is the quantity decisions require—the effect at the declared trial-eligible target population—and is what we report throughout. The movement from the published average to $\Delta^*$ separates cleanly into two parts, with three reference points: the *naïve published effect*; the *de-selected effect at the published control composition*, which removes selection but is still evaluated at the control mix the published trials embody; and $\Delta^*$, that de-selected effect re-expressed at the reference. The de-selected effect differs from $\Delta^*$ by the *target shift*, which equals $\beta_C(C_{pub} - \mu_C^*)$ when the published and trial-eligible covariate distributions coincide and otherwise also carries a covariate-composition term $\beta_W(\mu_{W,pub} - \mu_W^*)$; the naïve effect differs from $\Delta^*$ by the *sum* of that target shift and the selection correction, so which part is bias and which is a change of target must be read separately (Fig. 3).

What is identified from the published data alone is the bias-corrected outcome model and the *relative* selection profile; the *absolute* publication rate is not, and is supplied by registry calibration, which also falsifies the selection model where registry coverage permits. Because relevance is essential, we report a first-stage instrument-strength statistic $F$ for every application, computed *before* any correction, that flags the weak-instrument case. Because the control-arm mean enters the standardized effect itself, we guard against mechanical artifacts: a randomized-trial-cluster bootstrap, a multi-arm covariance that shares control arms across contrasts, and a re-expression on the native (unstandardized) outcome scale each leave the conclusions unchanged (Supplementary Materials).

> **Box 1 | What the external anchor identifies. Identified (published data + anchor):** the bias-corrected outcome model, the *relative* selection profile, and the target effect $\Delta^*$ at the declared population. **Not identified from published data alone:** the *absolute* publication/retrievability rate. **Supplied by registry calibration:** the selection intercept (the absolute publication level), which also falsifies the selection model where coverage permits. **Required for a publication-selection reading:** the external anchor approximates the completed-trial control distribution before publication; otherwise the selection component is selection of *published* evidence relative to the declared target. **Falsification / sensitivity:** the first-stage $F$ (relevance), the anchor-sensitivity curve, and a Copas-type direct $C \to R$ sensitivity.

> **Box 2 | The estimator in five lines. Input:** per-trial ($Y_k$, $SE_k$, $C_k$, $n_{C,k}$, $W_k$); external reference mean/variance of the untreated outcome; registry coverage rate. **Assumption checks:** compute first-stage $F$ (relevance); report it before correcting. If $F$ is small, stop and report nothing. **Fit:** maximize the published-only joint likelihood of ($Y_k$, $C_k$) under a probit selection function, with the marginal publication probability by Gauss–Hermite quadrature and the registry rate as a soft constraint. **Output:** $\Delta^*$ at the declared

target, with a randomized-trial-cluster bootstrap interval; a Copas-type exclusion-sensitivity curve; an anchor-sensitivity curve. **Report:** $\Delta^*$ alongside conventional estimates and the diagnostics, never the point estimate alone. *Interpretation:* the selection component is publication selection insofar as the external reference matches the completed-trial control distribution before publication; otherwise it is selection of published evidence relative to the declared target.

## A strict holdout benchmarks the correction against an observable all-trials benchmark

The answer in a publication-bias problem is normally unobservable. The antidepressant network meta-analysis of Cipriani *et al.* (*14*) is the exception: it assembled both published and recovered-unpublished trials—including regulatory and sponsor-held studies—on a common scale (HAM-D), and recorded each trial's publication status and source. This makes it a rare ground-truth setting in which a publication-selection correction can be validated out of sample. Treating the full published-plus-unpublished set as the benchmark gives an *observable all-trials benchmark* of 0.248 (HAM-D standardized mean difference; a plain random-effects recomputation over the 83 contrasts gives 0.285). On the published-only subset ($K$ = 71 contrasts), naïve pooling gives 0.31 (95% CI 0.25, 0.37; ≈ 2.5 HAM-D points), with significant funnel asymmetry (Egger $t$ = 3.9, P < 0.001); the instrument is very strong ($F$ = 56).

We put the method to the strictest test the data allow (Fig. 2). The 12 recovered-unpublished contrasts are removed from the analysis *entirely*—from both the estimation sample and the anchor—and the estimator is re-fit on the 71 published contrasts alone, so the hidden trials enter nowhere except the held-out benchmark. The published-only fit returns $\Delta^*$ = 0.22 (95% CI 0.15, 0.30; ≈ 1.8 HAM-D points). That interval contains the all-trials benchmark, and the point estimate lies exactly where it should: *below* the published naïve mean (0.36) and *above* the never-seen recovered-unpublished mean (0.19). The correction moves the published average toward the hidden evidence without ever observing it. What is recovered out of sample is the *aggregate* target estimand—the inverse-selection-weighted mean; the method does not claim to predict individual suppressed trials, since the recovered-unpublished set is the retrievable subset of all suppressed trials and the fitted selection slope is estimated near its boundary (Supplementary Materials).

Two further properties make this a positive control rather than a single lucky recovery. It is *specific*: applied to the full $K$ = 83 set with the registry constraint relaxed it returns 0.246, manufacturing essentially no correction where a literature is already near-complete. And it isolates the source of identification: conventional precision correctors miss in *opposite* directions in this same literature—PET drives the estimate to near zero (0.03) and p-uniform* degenerates, while naïve pooling and p-curve barely move (0.31, 0.35)—because they can read only funnel or p-value geometry, whereas the control-level anchor supplies additional identifying information unavailable to funnel- or p-value-only methods. We therefore frame the antidepressant literature as a benchmark recovery and calibration, not as proof that the corrections transfer unconditionally; each application below is gated by its own diagnostic. One point of scope: this benchmark exercises the selection-correction mechanism under a *near-complete trial-control* reference—the anchor derives from published or recovered control arms—whereas the insomnia and diabetes applications exercise the *fully external population-anchor* use case, in which the reference is an independent epidemiological distribution.

## Target-population effects in two clinical literatures

With the method calibrated against a known benchmark, we apply it where the answer is not observable but the anchor is credible, reporting effects on both the standardized and the native clinical scale (Fig. 3).

**Insomnia.** For pharmacological treatment of chronic insomnia, the trial-eligible reference for total sleep time (TST) is documented in large epidemiological surveys, reweighted to trial-eligibility composition (reference SD ≈ 50 min). Re-extracted at contrast level from the systematic review of De Crescenzo *et al.* (*15*) ($K$ = 63), naïve pooling gives 0.34 (95% CI 0.26, 0.41; ≈ +17 min). The selection correction at the published composition takes this to 0.15 (≈ +7.5 min); the target shift to the reference takes it to $\Delta^*$ = -0.07 (95% CI -0.23, 0.05; ≈ -3.5 min), against a minimal clinically important difference for total sleep time commonly placed at 20–30 minutes (Supplementary Materials). The first stage is strong ($F$ = 25.7; instrument–effect correlation -0.55): trials enrolling more severe insomnia (shorter placebo-arm sleep) report larger standardized effects, making the short projection from the published composition (346 min) to the reference (370 min) credible. The estimate is centered near zero and, in the declared trial-eligible target, rules out the published-trial average as a clinically important gain. Sensitivity to the choice and measurement of the reference—survey versus instrument-based sleep—is reported in the Supplementary Materials.

**Type-2 diabetes.** For the placebo-controlled HbA1c literature analyzed (*16*)—arm-level HbA1c from 174 trials (450 arms, 86,940 participants)—HbA1c is an objective biomarker whose trial-eligible distribution is documented epidemiologically (reference SD ≈ 0.7%), so the anchor rests on firmer ground than any psychometric scale. The literature shows pronounced funnel asymmetry (Egger $t$ = 9.6, $P < 10^{-4}$) and a naïve effect of 0.71 (≈ 0.50 HbA1c percentage points). The instrument is of *moderate* strength ($F$ = 7.2; "moderate" is a method-specific reading calibrated by matched simulation, not the conventional IV cutoff)—a deliberate test away from the strong-instrument regime—so we present this as an objective-outcome generalization, not a validation. The target-population aggregate is $\Delta^*$ ≈ 0.40 (95% CI 0.16, 0.55; ≈ 0.28 percentage points)—about forty percent below the published aggregate, with the wider interval that moderate instrument strength implies; a registry-constrained, model-averaged variant gives a concordant 0.35; we treat 0.40 as the primary single-specification estimate and 0.35 as the registry-constrained and baseline-instrument sensitivity value, the conclusion being the shared range 0.35–0.40. Because endpoint HbA1c can reflect both baseline severity and placebo-arm drift (adherence, trial duration, rescue therapy), we repeated the analysis using the placebo arm's *baseline* HbA1c as the instrument and obtained the same registry-constrained estimate (0.35); the correction is also stable across drug classes (leave-one-class-out $\Delta^* \in [0.32, 0.39]$). It is an aggregate generalization for this literature, not a clinical verdict on any drug class (Supplementary Materials).

These corrections are smaller than the published averages, but the direction and magnitude are set by *target and selection together*, not by a blanket downward rule. In a near-complete, funnel-symmetric antihypertensive blood-pressure literature—72 placebo-controlled ACE-inhibitor trials, with the placebo arm's achieved SBP as the instrument—Egger is null and PET/PEESE sit on the naïve −6.9 mmHg reduction, so the publication-selection correction is essentially nil; yet re-expressing at the reference control level (achieved SBP ≈ 150 mmHg) moves the effect *away* from zero, to $\Delta^*$ = −11.6 mmHg (95% CI −17.1, −6.7). Here, uniquely, the corrected effect is *larger* in magnitude than the naïve aggregate—the recovered selection slope is positive, so anchoring *increases* the implied reduction—showing that $\Delta^*$ re-expresses an effect at a declared target rather than shrinking it toward zero (Supplementary Materials).

## The method reports when it cannot be used

A central design goal is that the method announce its own inapplicability rather than silently return a biased number, and the same diagnostics that gate the applications above also rule literatures out (Table 1). The clearest case sits inside the insomnia trials themselves: their self-rated *sleep-quality* outcome is measured on 21 heterogeneous scales with no common epidemiological referent, its instrument is *weak* ($F$ = 2.4 versus 25.7 for TST in minutes), and we therefore report no corrected effect for it—the method refuses to answer when the anchor is not scientifically meaningful. Conversely, the method manufactures no spurious selection correction where none is warranted: the antidepressant full set returns the benchmark, and a funnel-symmetric blood-pressure literature shows the selection part vanishing while the target shift remains a separate re-expression at the reference (Supplementary Materials).

These cases instantiate a four-regime applicability map (Supplementary Materials). With a *valid anchor and a strong first stage*, the estimator is near-unbiased and near-nominally calibrated. With a *valid anchor but a weak first stage*, estimates become imprecise and the first-stage $F$ declines them in advance (sleep quality, $F$ = 2.4). With a *misspecified anchor*, bias grows approximately in proportion to the error in the anchor *mean* while staying insensitive to its variance, so the anchor-sensitivity curve bounds it. With an *exclusion violation*, the Copas-type curve bounds the residual control-level→publication coupling. The method is trustworthy in the first regime, and each of the three ways it degrades is flagged by a diagnostic available before or alongside the estimate.

Simulations confirm these operating characteristics (Supplementary Materials). The estimator is *calibration-oriented*: given a credible anchor and adequate relevance, it is near-unbiased and near-nominally calibrated (across a matched probit, a misspecified logistic link, and a non-smooth p-value step: bias +0.006, -0.011, +0.017; coverage 0.88, 0.96, 0.93). In this simplest cell a precision corrector (PET) is comparably calibrated, because it absorbs the small-study slope; the method's advantage is not lower error there but identification in the settings that demand the external anchor's content—target transport, study-size confounding, and continuous effect-dependent selection—where PET and the other conventional correctors reproduce the selected mean and are biased by 0.2 or more. The contribution is the external source of identification, not a contest among estimators that all use only the published record.

> **Box 3 | Before trusting a reference-anchored estimate, check five things.** 1. **Credible anchor** — is there an external, trial-eligible distribution of the untreated outcome that owes nothing to the published trials? 2. **First-stage relevance** — is $F$ large enough (computed before correcting) that the control-arm instrument actually moves with the effect? 3. **Exclusion sensitivity** — does the Copas-type curve show no qualitative reversal as control level is allowed to act on publication directly? 4. **Registry calibration** — is the absolute publication level pinned to a registry denominator (retrievable results), with the selection model falsified where coverage permits? 5. **Target interpretation** — is the declared reference population the one the decision concerns, and is the effect reported on a native scale a clinician can read?

## Discussion

Reference-anchored meta-analysis reframes evidence synthesis from "correcting the published average" to "choosing a target population and identifying selection with the same external anchor." The shift matters because the two problems it addresses are usually conflated: a pooled effect that looks inflated may reflect publication selection, target mismatch, or both, and only by separating them can a decision-maker read off the quantity that applies to patients. Where the answer is observable, a strict holdout recovers it; where it is not, the method reports target-population effects that are smaller than the published averages for the endpoints and literatures analyzed—compatible with no clinically important gain for total sleep time in the declared insomnia target, and about forty percent smaller for the placebo-controlled HbA1c literature under the external-anchor specification—while declining to answer where the anchor is weak. Published-trial averages can overstate target-population effects when selection and target mismatch align; whether they do, and by how much, is exactly what the external anchor makes estimable. These are re-estimates of what the assembled evidence implies for a declared target population, not efficacy verdicts on any treatment, dose, or class: the method changes what an evidence synthesis estimates, not what any individual drug does.

The contribution is identification, not a claim that an estimator using more information must beat estimators that use less. Published-only methods can exploit only funnel or p-value geometry, both uninformative when selection depends continuously on the effect; an external reference distribution, independent of the published record, supplies identifying information unavailable to published-only funnel- or p-value geometry, and that external information *is* the contribution. The closest existing approach, robust Bayesian model averaging (*9–11*), regularizes and averages over published-only selection models, whereas our approach adds an external source of identifying variation; the two are complementary, and a model-averaged variant of our selection function recovers its calibration when the truth lies in its class and improves on it when it does not (Supplementary Materials). The framework also connects to proximal causal inference and to data fusion (*17–19*): the control-arm mean is a treatment-side proxy whose external anchoring is what makes the otherwise-unobserved instrument usable.

Several limitations bound the method. It requires that editorial decisions plausibly depend on the standardized effect and observable study characteristics but not directly on the absolute control level given those quantities; this could fail if control severity acts on publication through other pathways (perceived clinical importance, sponsor strategy, drug class, or era), so for each application we enumerate the covariates that block such pathways and report a Copas-type sensitivity analysis—on an interpretable scale, spanning direct control-level effects on publication odds of up to roughly twofold per standard deviation—which shows no qualitative reversal in any literature analyzed (Supplementary Materials). It requires a credible external reference; where that reference is a heterogeneous clinical scale rather than an objective or norm-referenced measure, the anchor is weaker and the first-stage diagnostic should gate its use. It addresses publication selection across trials, not selective *outcome* reporting within them. Within these bounds the method is, above all, *auditable*: it states what is identified (the relative selection profile and the bias-corrected effect at a declared target) and what is not (the absolute publication rate, which needs registry calibration), and its diagnostic makes clear, in advance, when the data can and cannot support a correction. The broader implication for evidence-based medicine is concrete: a pooled effect is not a property of a treatment alone but of a treatment evaluated at a particular, often unstated, control population—and making that population explicit, and anchoring it externally, can change what the evidence says.

## Materials and Methods

**Estimation in brief.** The model is fit by penalized maximum likelihood of the published-only joint density of the reported effect and the control-arm instrument, with the marginal publication probability evaluated by 24-point Gauss–Hermite quadrature and the registry-calibrated rate entering as a soft constraint; a no-selection component with model-averaging weights stabilizes the estimator when the selection signal is weak. The control-arm mean is treated as an error-prone measurement (its sampling variance $\sigma_C^2/n_{C,k}$ carried through), and robustness to shared control arms in multi-arm trials, to trial-level clustering, and to native-scale re-expression is verified by sensitivity analysis. Each fit uses four deterministic multi-starts and reached a common optimum in every application. The primary likelihood propagates the sampling error in each control-arm mean (errors-in-variables); shared control arms in multi-arm trials are handled in a multi-arm covariance sensitivity analysis. All interval estimates are bootstrap percentile intervals resampled at the randomized-trial (cluster) level—so that contrasts sharing a control arm are resampled together—and also resample the registry-calibration step. We assess frequentist calibration by simulation and by the strict-holdout positive control rather than by asymptotic theory.

**Data and code availability.** The antidepressant dataset of Cipriani *et al.* (*14*) is openly available; the type-2 diabetes HbA1c data of Kuss *et al.* (*16*) are available under CC-BY-4.0; the insomnia contrasts are re-extracted from the open systematic review of De Crescenzo *et al.* (*15*). All code, simulation scripts, and bundled data reproduce every headline result, simulation, and figure from a single command, including the strict-holdout experiment. The replication package accompanies this preprint (and is available from the authors); a permanent, citable archive (Zenodo DOI) will be deposited upon publication.

**Acknowledgments:** The authors thank seminar participants at Keio University and RIKEN AIP for valuable discussions on identification strategies in meta-analytic contexts. **Funding:** This work was supported by JSPS KAKENHI Grant Number 25K21651. **Author contributions:** T.H. designed research; K.D., R.N., and T.H. developed the methodology and performed research; K.D. and R.N. analyzed data; K.D., R.N., and T.H. wrote the paper. **Competing interests:** The authors declare no competing interests. **Data and materials availability:** All data and code are in the replication package accompanying this article.

## List of Supplementary Materials

# Figures

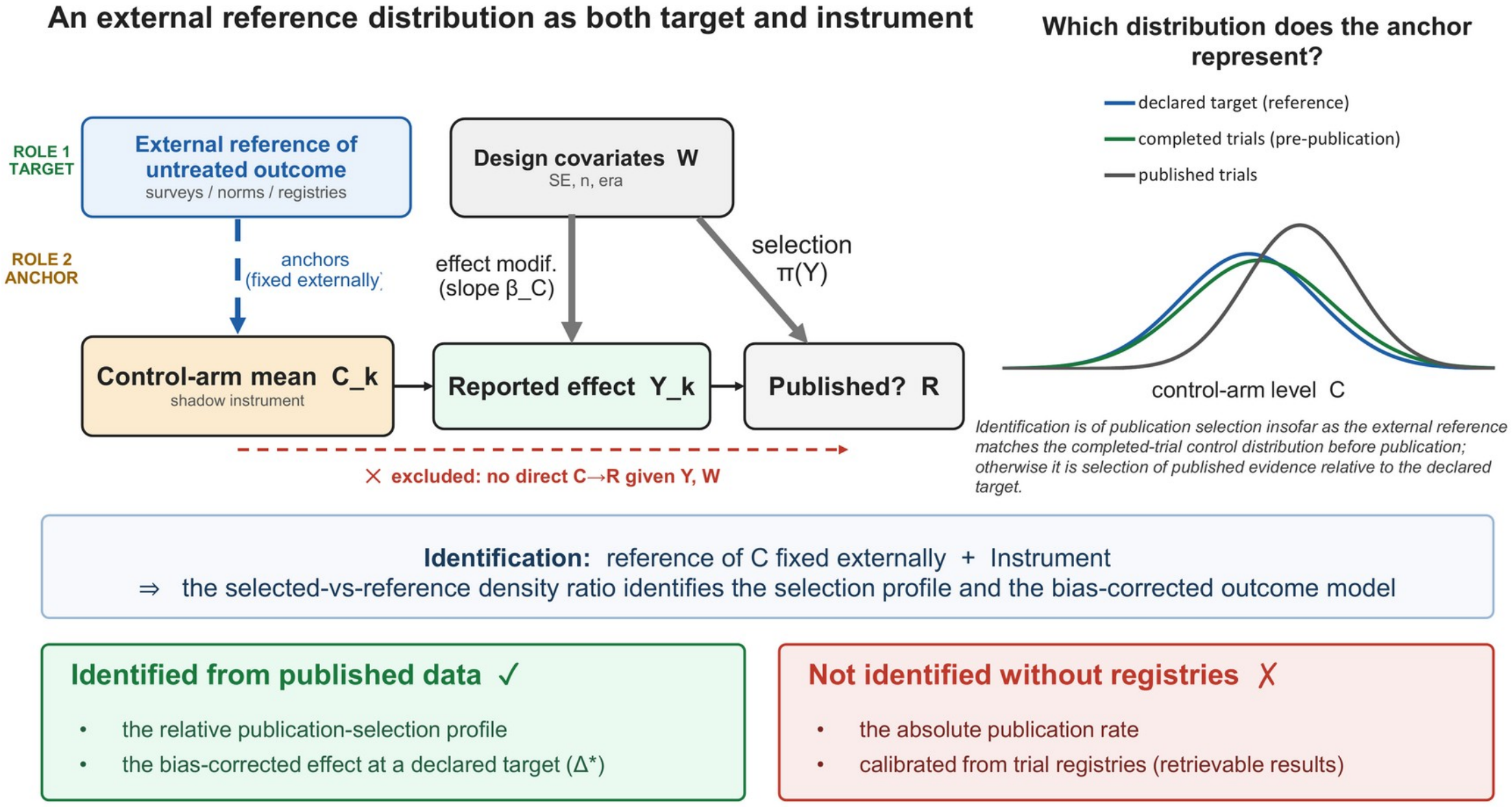

**Fig. 1. An external reference distribution as both target and instrument.** Concept and causal diagram for the dual role of the external control-outcome reference. The control-arm mean $C_k$ is a shadow instrument: associated with the reported effect $Y_k$ through baseline-severity effect modification, but with no direct path to publication $R$ given the reported effect and design covariates. The one-line identification logic: with the reference distribution of $C$ fixed externally and $R \perp C \mid Y, W$, SE, the selected-versus-reference density ratio identifies the selection profile and the bias-corrected outcome model; the *relative* profile and the effect at a declared target are identified from published data, while the *absolute* publication rate is not and is calibrated from registries. As the target, the reference fixes the control level at which the effect is reported (the estimand $\Delta^*$). The figure names the three control-level distributions the analysis distinguishes—completed trials (before publication), published trials, and the declared target (external reference); the selection component is publication selection insofar as the reference matches the completed-trial distribution, and is otherwise selection of published evidence relative to the declared target.

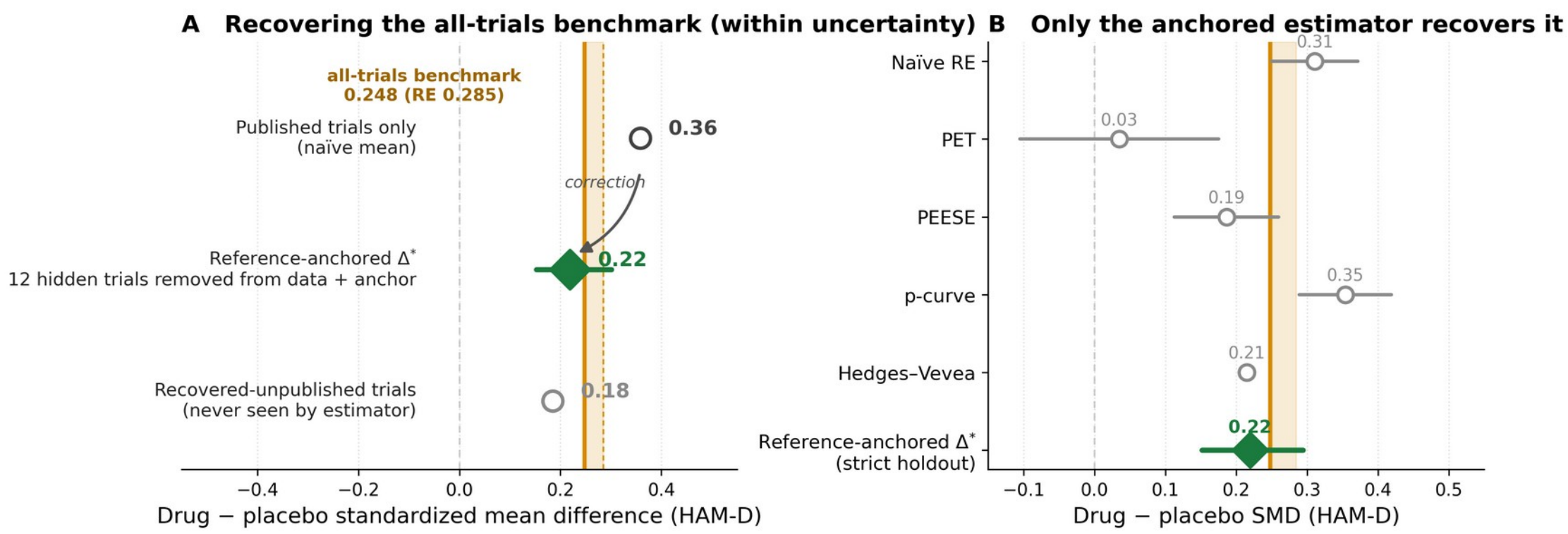

**Fig. 2. A strict holdout benchmarks the correction against an observable all-trials benchmark.** Antidepressant positive control (Cipriani *et al.*, *14*). **(A)** The 12 recovered-unpublished contrasts are removed from *both* the estimation sample and the anchor; the estimator is re-fit on the 71 published contrasts alone. The published-only fit ($\Delta^*$ = 0.22, 95% CI shown) contains the all-trials benchmark (amber band, 0.248; the random-effects recomputation 0.285 marked) and falls between the published naïve mean (0.36) and the never-seen recovered-unpublished mean (0.19, open marker): the correction moves the published average toward hidden evidence the estimator never sees—a rare out-of-sample validation that recovers the aggregate target estimand, not a prediction of which individual trials were suppressed. **(B)** On the same published-only subset, only the anchored estimator recovers the benchmark; precision and p-value correctors miss in opposite directions because they read funnel or p-value geometry rather than the control-level anchor.

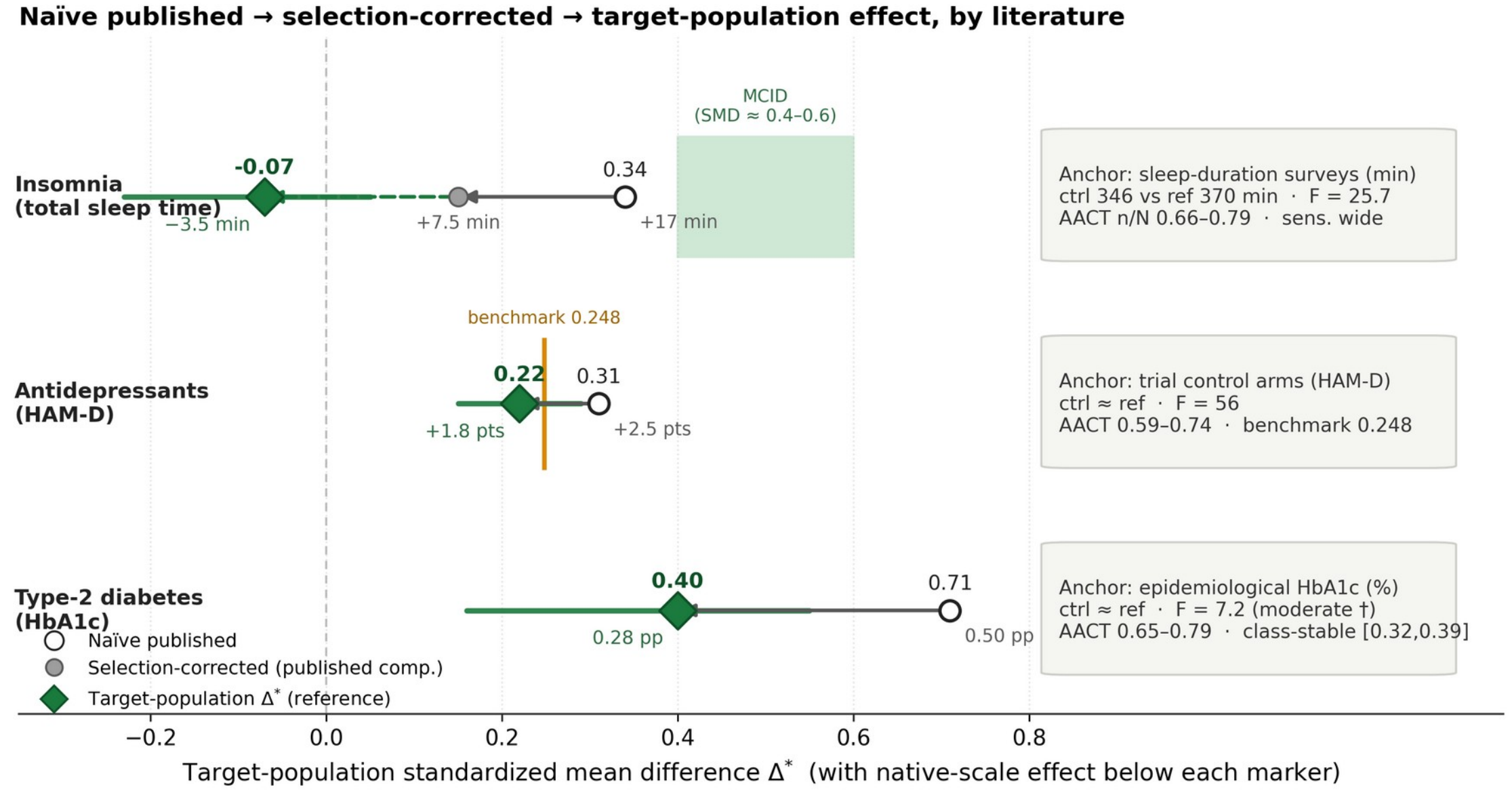

**Fig. 3. Target-explicit decomposition across three literatures, with native units and anchor audit.** Each lane shows the naïve published effect (open circle) reduced first by a publication-selection correction at the published control composition (grey, solid arrow) and then by a target shift to the trial-eligible reference (green diamond, dashed arrow), with 95% CI; the right axis of each lane gives the native-scale effect (sleep minutes; HAM-D points; HbA1c percentage points). The de-selected effect re-expressed at the reference equals the target shift, $\beta_C(C_{pub} - \mu_C^*)$ when covariate composition is matched between the published and target populations; the naïve-to-$\Delta^*$ gap is the sum of both steps. Each lane carries an anchor badge: external source, published-versus-reference control level, first-stage $F$, and the anchor-sensitivity range. For insomnia the published trials are more severe than the reference, so the target shift is large and negative and $\Delta^*$ falls below the minimal clinically important difference (shaded band); for the antidepressant and diabetes literatures the published composition is near the reference, so the shift is negligible. In the antidepressant lane the all-trials benchmark (amber line, 0.248) lies within the CI of $\Delta^*$. † marks the moderate-instrument diabetes case.

## Table 1

**Table 1. Applicability diagnostics: where the method does and does not apply.** For each literature, the externally specified anchor and the *a priori* diagnostics that gate the correction. The first-stage $F$ is computed before any correction; the method reports a corrected effect only where the anchor is credible and $F$ is adequate. Self-rated sleep quality—measured inside the same insomnia trials—is shown as a case the method declines.

| Literature / endpoint | External anchor (source) | First-stage $F$ | Registry calibration (fit) | AACT audit range | Status |
|---|---|---|---|---|---|
| Insomnia — total sleep time | Sleep-duration surveys (min) | 25.7 (strong) | 0.60 | 0.66–0.79 | **Reported:** $\Delta^*$ = −0.07 (−3.5 min) |
| Antidepressa | Combined / | 56 (strong) | 0.79 ‡ | 0.59–0.74 | **Reported +** |

| Literature / endpoint | External anchor (source) | First-stage $F$ | Registry calibration (fit) | AACT audit range | Status |
|---|---|---|---|---|---|
| nts — HAM-D | published control arms | | | | **benchmarked:** $\Delta^{*}$ = 0.22 (benchmark 0.248) |
| Type-2 diabetes — HbA1c | Epidemiological HbA1c (%) | 7.2 (moderate †) | 0.75 | 0.65–0.79 | **Reported (generalization):** $\Delta^{*}$ = 0.40 (0.28 pp); sens. 0.35–0.40 |
| Insomnia — sleep quality | none (21 heterogeneous scales) | 2.4 (weak) | — | — | **Declined** (weak instrument) |
| Antihypertensives — blood pressure | Achieved SBP norms (near-complete) | (SI) | — | — | **Specificity:** selection ≈ 0; re-expression away from zero (SI) |

Table note. The reported estimand is $\Delta^{*}$ at the trial-eligible reference. Registry calibration lists the prespecified fit value; the AACT audit range is an independent retrievable-result check, not an identical fitting input. ‡ For antidepressants the fit (0.79) exceeds the AACT range because that range is a lower bound—regulatory and sponsor-recovered studies, which enter the benchmark, are not captured by ClinicalTrials.gov linkage. † $F$ = 7.2 is a method-specific reading calibrated by matched simulation, not the conventional IV cutoff; the diabetes application is an objective-outcome generalization for the literature analyzed, not a clinical verdict on any drug class. For antidepressants the observable all-trials benchmark is 0.248, recovered out of sample by the strict holdout ($\Delta^{*}$ = 0.22, 95% CI 0.15, 0.30). Published-vs-reference control levels appear in Fig. 3, and the exclusion (direct control-level → publication) sensitivity in the Supplementary Materials. Full per-literature decompositions and native-scale conversions are in the Supplementary Materials; a condensed anchor audit is provided in Table 2, and the full construction (calendar years, inclusion/exclusion, reweighting) in Table M5. The eleven-method comparison and per-stratum registry counts (by era and registration status), the audit script, and falsification checks are in the Supplementary Materials (Tables M4, S16). The minimal clinically important difference for total sleep time (20–30 min) is sourced in the Supplementary Materials.

## Table 2

**Table 2. Anchor audit: construction of each external reference.** The external anchor is the method's identifying input, so its provenance is summarized here and given in full in Table M5. Objective or norm-referenced anchors (sleep duration, HbA1c, achieved blood pressure) rest on firmer ground than heterogeneous clinical scales; the antidepressant anchor is built internally

from the near-complete combined control arms and agrees with the published-only route. “Baseline/endpoint” records whether the reference is the untreated baseline or the achieved control-arm endpoint.

| Literature / endpoint | Anchor source & trial-eligible population | Measurement (baseline/endpoint) | Reference mean (SD) | Anchor-sensitivity |
|---|---|---|---|---|
| Insomnia — total sleep time | Sleep-duration surveys, adults with insomnia | Survey / actigraphy; endpoint | 370 min (50) | over documented survey/device range |
| Antidepressants — HAM-D | Pooled control arms, combined published + recovered (internal) | Clinician-rated HAM-D; endpoint | 14.7 / 17.4 (≈8) | published-only = combined |
| Type-2 diabetes — HbA1c | HbA1c in sub-optimally controlled type-2 diabetes | Laboratory HbA1c; baseline & endpoint | 8.0% (0.7) | range 0.35–0.40 |
| Insomnia — sleep quality | none (21 heterogeneous scales; no referent) | no common scale | — | — |
| Antihypertensives — blood pressure | Achieved SBP norms (near-complete) | Office SBP; endpoint | 150 mmHg (18) | specificity case |

# Supplementary Materials for

## External anchors reveal target-population effects hidden by published clinical-trial evidence

**Kasumi Dan, Ryoya Nakano, Takahiro Hoshino**

Corresponding author. Email: hoshino@econ.keio.ac.jp

**This PDF file includes:**

Materials and Methods (M1 to M6) Supplementary Text (S1 to S17) Figs. S1 to S5 Tables S1 to S16 References

---

# Materials and Methods

The main text states the finding; this section gives the model, the identification result, the estimator, and the data in the detail needed to reproduce every reported number from the accompanying replication package. Sections M1–M6 are self-contained; the Supplementary Text (S1–S17) gives proofs, the full simulation study, sensitivity analyses, and additional applications.

## M1. Study-level model and target estimand

For each trial $k$ we observe a standardized treatment effect $Y_k$ (a drug-versus-control standardized mean difference), its standard error $SE_k$, the control-arm outcome mean $C_k = \hat{\mu}_{0,k}$ and control-arm size $n_{C,k}$, and design covariates $W_k$ (a subset $U_k$ of which may enter selection). The structural outcome model lets the true effect depend on the control-arm severity,

$$Y_k = \alpha + \beta_C C_k + \beta_W^\top W_k + u_k, \quad u_k \sim N(0, \tau_\delta^2 + \mathrm{SE}_k^2),$$

so the slope $\beta_C$—baseline-severity *effect modification*—is the channel that makes the control mean informative about the effect. We write $T_k = Y_k/SE_k$ for the trial's test statistic, the argument of publication selection.

An external source supplies the mean $\mu_C^{pop}$ and variance $\sigma_C^2$ of the untreated outcome in the *trial-eligible* population. Because $C_k$ is a control-arm sample mean of size $n_{C,k}$, a central-limit bridge maps these individual-level moments to the study-level distribution of the observed control mean,

$$C_k \mid W_k, n_{C,k} \sim N(\mu_C^{\mathrm{pop}}, \tau_0^2 + \sigma_C^2/n_{C,k}), \qquad \text{(M1)}$$

where $\tau_0^2$ is between-study dispersion of the latent control level. We standardize the instrument as $Z_k = (C_k - \mu_C^{pop})/\sigma_C$.

The reported estimand is the bias-corrected effect at the reference control level, averaged over the trial-eligible covariate distribution,

$$\Delta^* = \int \mathbb{E}_\theta[Y \mid C = c, U = u] m^*(c \mid u) dP^*(u) dc, \qquad \text{(M2)}$$

where $m^*$ is the reference (target) control distribution. We distinguish three control distributions: $m_0$, the distribution that generated the *completed* trials *before publication*; $m_A$, the analyst-supplied anchor; and $m^*$, the target population. In the main applications we set the anchor equal to the declared target, $m_A=m^*$, **by design**. Interpreting the selected-versus-reference component specifically as *publication* selection additionally requires $m_0 \approx m_A$; when this approximation is doubtful—for example if more severe cases are not only published more readily but also *enrolled* more readily—the estimate should be read as selection of the published evidence relative to the declared target, with the target shift reported separately. The framework also permits $m^*$ to differ from $m_A$, in which case movement of $\Delta^*$ with $m^*$ is a deliberate change of target (transport), not bias (S5). The naïve published pooled effect differs from $\Delta^*$ by exactly $\beta_C (C_{pub}-\mu_C^*)$, decomposing the gap into a publication-selection part and a target-population part.

## M2. Identification: exclusion restriction, completeness, and what registries add

Three assumptions identify the model. **Assumption 1 (exclusion).** Publication depends on the reported effect and observable covariates but not on the control level given those: $C \perp R \mid Y, U$, i.e. $\pi(Y,U;\gamma)=\Pr(R=1 \mid Y,U)$ does not depend on $C$. This is the shadow-instrument restriction: a trial's absolute control severity affects publication only through the reported contrast. **Assumption 2 (external anchor).** The control-distribution $p(C \mid U)$ is known from the external reference (M1), not estimated from the published sample. **Assumption 3 (completeness/relevance).** For each $U$, the outcome family $\{p(Y \mid C=c,U;\theta): c \in \text{supp}(C \mid U)\}$ is bounded-complete; in the linear-Gaussian case this reduces to $\beta_C \neq 0$.

**Theorem 1 (informal).** Under Assumptions 1–3, the bias-corrected outcome model ($\theta$, hence $\Delta^*$) and the *relative* selection profile $\gamma_{rel}$ are identified from the published-only joint distribution of $(Y,C,U)$. The *absolute* publication level—the $U$-specific intercept $a(U) \equiv \Pr(R=1 \mid U)$, equivalently the link intercept $\gamma_0$—is **not** identified from published-only data and requires auxiliary calibration. The proof (S1) shows that taking ratios of the observed selected density at two control levels $c,c'$ cancels the selection function and its normalizer, leaving the outcome-density ratio, which identifies $\theta$ by completeness; the relative selection profile then follows by inversion, with only its absolute level unidentified.

The intuition is that the externally fixed control distribution supplies exogenous variation in $C$ that traces the regression $\alpha+\beta_C C$ *independently of the selection mechanism*, because the instrument's distribution is supplied from outside the selected sample rather than estimated within it. This is what classical nonresponse-IV and proximal methods cannot do here: the instrument $C$ is itself unobserved for the unpublished trials (S9).

## M3. Estimation: penalized maximum likelihood with a registry-calibrated constraint

The headline estimates are computed by **penalized maximum likelihood**, not by Markov-chain Monte Carlo; the estimator uses `scipy` optimization with Gauss–Hermite quadrature and multi-start, and intervals are obtained by a randomized-trial-cluster bootstrap. For published study *k* the likelihood contribution is the product of three factors:

1. an **outcome density** $N(Y_k;\delta+\beta_C Z_k,\tau_\delta^2+\mathrm{SE}_k^2+\beta_C^2 v_{0,k})$, where $v_{0,k}=\sigma_C^2/n_{C,k}$ is the control-mean sampling variance entering as an errors-in-variables term (S8);
2. a **selection factor** $\Phi(\gamma_0+\gamma_1 T_k+\gamma_2 T_k^2)$, a probit of the test statistic (the linear specification sets $\gamma_2=0$; the quadratic estimates it; S15); and
3. the **reciprocal marginal publication probability** $[\int\Phi(\cdot)p(Y\,|\,Z_k)dY]^{-1}$, evaluated by 24-point Gauss–Hermite quadrature.

The registry-calibrated absolute publication rate enters as a soft quadratic penalty on the mean marginal publication probability (M4). The free parameters are $(\delta,\tau_\delta,\gamma_0,\gamma_1,\gamma_2,\beta_C)$; the reported estimand $\Delta^*$ is the fitted effect at the reference level Z = 0. A no-selection component with model-averaging weights stabilizes the estimator when the selection signal is weak, without blunting the correction when selection is genuinely present (S3, S5).

**Optimization.** Each fit is run from four dispersed deterministic starting points under box constraints (L-BFGS-B). In every application all four starts converged to a common optimum, with the reference-effect estimate agreeing across optimal starts to within $2\times10^{-6}$ (Table S2); there are no boundary or multimodality pathologies of the kind that affect classical step-function selection models.

**Intervals.** Reported 95% intervals are randomized-trial-cluster bootstrap percentile intervals that *also* resample the registry-calibration step, so registry-rate uncertainty is propagated (M4, S7). The resampling unit is the randomized trial, not the contrast: all contrasts sharing a randomized control arm (multi-arm trials) are resampled together as a single cluster, and the shared-control covariance is retained. We assess frequentist calibration by simulation (S3) and by the positive control (S12) rather than by appeal to asymptotic theory.

**First-stage diagnostic.** Because relevance is essential and is checkable before any correction, we report, for every application, the first-stage instrument-strength statistic *F* for the regression of *Y* on the standardized control mean *Z* (with design covariates), computed *a priori*. A weak first stage (*F* small) is treated as a signal that the data cannot support the correction, and no corrected effect is reported (S11).

## M4. Registry calibration of the absolute publication rate

Theorem 1 leaves the absolute publication level unidentified from published data. We calibrate it from trial registries. For registry stratum *s* we model the retrievable-result count

$M_s\sim\mathrm{Binomial}(N_s,q_s)$ with $q_s=a_s\mathbb{E}_{\theta,\gamma}[\rho_\gamma(Y,U)\,|\,U\in s]$, where $N_s$ is the number of registered trials in the stratum. The stratum rates $\{q_s\}$ enter the likelihood as the soft constraint of M3. The same registry data provide a *falsification* check: the model-implied publication rate in each stratum is compared against the observed retrievable-result fraction, and a stratum whose discrepancy exceeds its binomial standard error would reject the selection model. For the insomnia registry strata the implied and observed rates agree within sampling error (S7). We audit these rates directly from ClinicalTrials.gov through the AACT database. For each application a registry cohort is defined by condition MeSH together with the relevant drug set—glucose-lowering agents; the antidepressants of Cipriani *et al.*; hypnotics and sedatives—restricted to interventional phase 2–4 trials, completed, registered within the meta-analytic window. The denominator $N_s$ counts registered eligible trials; the numerator $M_s$ counts those with a *retrievable result*, operationalized as results posted to ClinicalTrials.gov **or** a linked result publication (the sponsor-designated result reference, with the broader auto-linked set as the upper variant). The audited coverage is 0.66–0.79 (insomnia, $N$=131), 0.59–0.74 (antidepressant, $N$=241), and 0.65–0.79 (type-2 diabetes, $N$=885), the range spanning the conservative and broad publication-linkage numerators. Regulatory-only results (Drugs@FDA, EMA EPAR) are not captured, so these are lower bounds—most consequentially for the antidepressant literature, whose combined set additionally draws on regulatory and sponsor-supplied unpublished sources. The reported fits use prespecified calibration values (0.60, 0.79, 0.75), compared with—but not identical to—the AACT audit: for insomnia the fitted value is *conservative* relative to the audited range; for antidepressants the AACT publication-linkage range is a *lower bound*, because regulatory and sponsor-recovered studies (which enter the benchmark) are not captured by ClinicalTrials.gov linkage; for diabetes the fitted value lies within the audited range. All conclusions were stable across the audited range (S6, S7). Per-stratum coverage rises with prospective registration and the post-FDAAA era (Table M4), confirming the era × registration-status stratification.

Table M4 reports the audited cohorts, the aggregate $N$/$n$/coverage per application, and the per-stratum breakdown; the counts ($N_s$,$M_s$) and the audit script (`registry_audit.py`, operating on the AACT `studies`, `calculated_values`, `browse_conditions`, `browse_interventions`, and `study_references` tables) are provided in the replication package.

**Table M4. Registry audit of the absolute publication rate (ClinicalTrials.gov / AACT).**

| Application | Cohort (condition × drug set) | Window | Registered (N) | Retrievable (n) | Coverage (n/N) |
|---|---|---|---|---|---|
| Insomnia TST | insomnia × hypnotics/sedatives | reg ≤2021 | 131 | 87–103 | 0.66–0.79 |
| Antidepressant HAM-D | major depression × Cipriani antidepressants | reg ≤2016 | 241 | 141–178 | 0.59–0.74 |
| Type-2 diabetes HbA1c | type-2 diabetes × glucose- | reg ≤2015 | 885 | 574–703 | 0.65–0.79 |

| Application | Cohort (condition × drug set) | Window | Registered (N) | Retrievable (n) | Coverage (n/N) |
|---|---|---|---|---|---|
| | lowering agents | | | | |

Retrievable *n* ranges from the sponsor-designated-publication numerator (lower) to the broader auto-linked-publication numerator (upper), each unioned with registry results posting; regulatory-only results are not captured (lower bound). Cohorts are interventional phase 2–4 completed trials. Per-application falsification: insomnia—model-implied and observed rates agree within binomial SE per stratum; antidepressant—regulatory sources raise *n* toward the near-complete benchmark; type-2 diabetes—leave-one-drug-class-out stable.

**Per-stratum coverage (era × registration status; retrievable = posting ∪ result publication, conservative numerator), *n*/*N*:**

| Application | ≤2007, retrospective | 2008–2016, prospective | 2008–2016, retrospective | 2017+, prospective |
|---|---|---|---|---|
| Insomnia | 15/27 (0.56) | 27/29 (0.93) | 18/33 (0.55) | 16/20 (0.80) |
| Antidepressant | 25/56 (0.45) | 72/96 (0.75) | 33/59 (0.56) | — |
| Type-2 diabetes | 69/147 (0.47) | 353/459 (0.77) | 99/191 (0.52) | — |

*Coverage rises with prospective registration and the post-FDAAA (2008+) era in every literature—the empirical basis for the era × registration-status strata. The rate enters the likelihood only as a soft constraint on the absolute publication level, not the relative selection profile, and the headline conclusions are insensitive to where in the audited range it sits (S6, S7).*

The calibration enters only the *absolute* publication level, not the relative selection profile, so the headline conclusions (which depend on the relative profile and the target shift) are insensitive to it; its effect on the intervals is propagated by resampling the registry step in the bootstrap (S7). "Retrievable result" denotes a registered trial whose outcome is obtainable from any source (publication, registry results posting, or regulatory filing), which is broader than journal publication and is the quantity the calibration targets.

## M5. Construction of the external reference distributions

Each application requires an external, trial-independent reference for the trial-eligible untreated outcome.

*Insomnia (total sleep time).* The reference for TST is taken from large epidemiological surveys of sleep duration in adults with insomnia, reweighted to the demographic composition of the trial-eligible population; the individual-level reference SD is ≈ 50 min, against which the published-composition control mean (346 min) and the reference (370 min) define the target shift. *Type-2 diabetes (HbA1c).* The reference is the documented distribution of HbA1c in adults with sub-optimally controlled type-2 diabetes, with individual-level mean and SD ≈ 8.0% and 0.7%; because HbA1c is an objective biomarker its trial-eligible distribution rests on firmer ground than any psychometric scale. *Antidepressant (HAM-D).* Uniquely, the anchor can be built *internally* from the pooled control-arm distribution of the combined published-plus-recovered-unpublished set, because that set is (near) complete; the external and internal routes give nearly identical estimates (S12). Each control mean is mapped to the study level through (M1) using the trial's recorded $n_{C,k}$.

Table M5 documents, for each main application, the reference's source, the trial-eligible population definition, the reweighting to trial eligibility, the measurement modality, and the individual-level mean and SD with their approximate uncertainty—so that the anchor is auditable rather than asserted.

**Table M5. Construction of the external reference distributions (source, population, reweighting).**

| Application | Source of reference | Trial-eligible population | Reweighting |
|---|---|---|---|
| Insomnia TST | Epidemiological sleep-duration surveys in adults with insomnia | Adults meeting trial inclusion (chronic insomnia, no major comorbidity) | Reweighted to trial age/sex composition |
| Antidepressant HAM-D | Pooled control-arm endpoint of the combined published + recovered set (internal); cross-checked against published-only controls | Adults with major depression meeting trial inclusion | Within-scale (HAM-D-17/21) |
| Type-2 diabetes HbA1c | Documented HbA1c distribution in adults with sub-optimally controlled type-2 diabetes | Adults meeting trial inclusion (elevated HbA1c) | Reweighted to trial baseline-HbA1c band |

**Table M5 (continued). Measurement, reference mean, and anchor uncertainty.**

| Application | Measurement | Mean (SD) | Anchor sensitivity / uncertainty |
|---|---|---|---|
| Insomnia TST | Survey / actigraphy sleep duration | 370 min (50 min) | anchor-sensitivity over documented survey/device range; sampling CI excludes anchor uncertainty |
| Antidepressant HAM-D | Clinician-rated HAM-D endpoint | 14.7 / 17.4 (≈8) | published-only and combined anchors |

| Application | Measurement | Mean (SD) | Anchor sensitivity / uncertainty |
|---|---|---|---|
| | | | give identical estimates |
| Type-2 diabetes HbA1c | Laboratory HbA1c (%) | 8.0% (0.7%) | anchor-sensitivity over trial-eligible HbA1c range (S12/S13) |

**Anchor uncertainty.** The reported intervals propagate registry-rate uncertainty (M4) and, for the reference mean, are accompanied by an explicit anchor-sensitivity analysis rather than treated as fixed. For insomnia, the reference-choice curve traces $\Delta^*$ as the anchor mean is varied across the documented range (the same curve serves as a transportability analysis; S5); for depression the corrected effect moves only within 0.29–0.30 as the BDI-II anchor is varied over $N(20–26,8–12)$ (S13). The simulation anchor-misspecification study (S4, S5) shows bias is approximately proportional to the anchor-*mean* error and insensitive to the anchor *variance*, so the sensitivity of $\Delta^*$ to the anchor is governed by, and bounded by, the first-stage slope $\beta_C$ times the plausible anchor-mean range.

**Native-scale conversions.** The standardized $\Delta^*$ is converted to native units by multiplying by the reference SD; this is a *conversion* of the reported estimate, not an independent native-scale refit, and gives the same clinical interpretation in each literature.

| Endpoint | $\Delta^*$ (SMD) | Reference SD | Native-scale $\Delta^*$ | Clinical referent |
|---|---:|---:|---:|---|
| Insomnia TST | −0.07 | 50 min | −3.5 min | MCID 20–30 min |
| Antidepressant HAM-D | 0.22 | ≈8 | 1.8 HAM-D points | all-trials benchmark 0.248 |
| Type-2 diabetes HbA1c | 0.40 | 0.7% | 0.28 percentage points | naïve 0.50 pp |

## M6. Data sources and extraction

*Antidepressants.* The open dataset of Cipriani *et al.* records each trial's publication status and data source (published article, regulatory submission, sponsor-provided unpublished study). Our analytic subset (HAM-D-17 and HAM-D-21 endpoint scores, signed SMD in [-1,3]) yields 83 drug-versus-placebo contrasts from 60 trials—71 published and 12 recovered-unpublished. The all-trials benchmark $\delta_{truth}$=0.248 is random-effects pooling over the full 83-contrast set. *Insomnia.* Contrast-level TST effects (K = 63) are re-extracted from the open systematic review of De Crescenzo *et al.*, with control-arm sleep duration as the instrument. *Type-2 diabetes.* Arm-level endpoint HbA1c from 174 placebo-controlled trials (450 arms, 86,940 participants) from Kuss *et al.* (CC-BY-4.0), re-pairing each placebo arm with its largest active arm. Additional applications (education, psychotherapy, blood pressure, exercise) are described in S13. All extraction scripts and bundled data reproduce every headline number from a single command.

---

# Supplementary Text

## S1. Proofs of Theorem 1 and Corollary 1

**Setup.** Let $(Y,C,U,R)$ have joint density $p(y,c,u,r)$; the analyst observes an i.i.d. sample of $(Y,C,U)$ conditional on R = 1. The model is specified by $p(C| U)$ known (Assumption 2); $p(Y| C,U;\theta)$, the target outcome distribution; and $\pi(Y,U;\gamma)=\Pr(R=1| Y,U)\in(0,1]$, the publication probability, which under Assumption 1 does not depend on $C$. Define $\Omega(U;\theta,\gamma)=\iint \pi(y,U;\gamma)p(y\mid c,U;\theta)p(c\mid U)dydc$. By Bayes' rule and Assumption 1 the selected joint density factors as

$$p(Y,C\mid U,R=1)=\frac{\pi(Y,U;\gamma)p(Y\mid C,U;\theta)p(C\mid U)}{\Omega(U;\theta,\gamma)}. \qquad \text{(S1)}$$

**Identification conditions.** (R1) $\theta\mapsto p(\cdot|\cdot,U;\theta)$ is one-to-one for a.e. $U$. (R2) the *relative* selection profile $\gamma_{rel}\mapsto\pi(y,u;\gamma)/\pi(y_0,u;\gamma)$ is one-to-one for a.e. $u$; no condition identifies the $U$-specific level $a(U)$. (R3) for each $U$, $\{p(Y| C=c,U;\theta):c\in\mathrm{supp}(C| U)\}$ is bounded-complete (Newey–Powell; D'Haultfœuille).

**Lemma S1 (identification of $\theta$).** Under Assumptions 1–3 and (R1), $\theta$ is identified from $p(Y,C| U,R=1)$. *Proof.* Define $\rho(y,c| U)=p(y,c| U,R=1)/p(c| U)$. By (S1), $\rho(y,c| U)=\pi(y,U;\gamma)\, p(y| c,U;\theta)/\Omega(U;\theta,\gamma)$. For two control levels $c,c'$,

$$\frac{\rho(y,c\mid U)}{\rho(y,c'\mid U)}=\frac{p(y\mid c,U;\theta)}{p(y\mid c',U;\theta)},$$

so $\pi$ and $\Omega$ cancel and the left side is a function of observables. If $\theta_1,\theta_2$ generate the same observed law, then $\partial_y\log p(y\mid c,U;\theta_1)-\partial_y\log p(y\mid c,U;\theta_2)$ is independent of $c$; combined with $\int p(y\mid c,U;\theta)dy=1$ this forces $p(\cdot| c,U;\theta_1)=p(\cdot| c,U;\theta_2)$, hence $\theta_1=\theta_2$ by (R1). □

**Lemma S2 (identification of $\gamma_{rel}$).** Given $\theta$ identified, integrating (S1) over $c$ gives $p(y| U,R=1)=\pi(y,U;\gamma)\, p(y| U;\theta)/\Omega(U;\theta,\gamma)$ with $p(y\mid U;\theta)=\int p(y\mid c,U;\theta)p(c\mid U)dc$ computable. Hence

$$\log\pi(y,U;\gamma)-\log\pi(y_0,U;\gamma)=\log\frac{p(y\mid U,R=1)}{p(y_0\mid U,R=1)}-\log\frac{p(y\mid U;\theta)}{p(y_0\mid U;\theta)}$$

is identified for any reference $y_0$, identifying $\gamma_{rel}$ by (R2). The constant $\Omega(U)$—equivalently $\gamma_0$, equivalently $\Pr(R=1| U)$—is not identified from published-only data. □

**Theorem 1 (formal).** Under Assumptions 1–3 and (R1)–(R2), $\theta$ and $\gamma_{rel}$ are identified from $p(Y,C,U|\,R=1)$; the level $a(U)$ is identified only after registry calibration (M4). Completeness (R3) is what is needed if the outcome model is treated nonparametrically rather than parametrically, generalizing Lemma S1 to nonparametric uniqueness in the spirit of nonparametric IV.

**Corollary 1 (linear-Gaussian).** With $Y\,|\,C,U \sim N(\alpha+\beta_C C+\beta_U^\top U, \sigma_Y^2)$, $\Pr(R=1|\,Y,U)=\Phi(\gamma_0+\gamma_1\,Y+\gamma_2$^⊤ U), and $C|\,U\sim\,N(\mu_C(U),\sigma_C^2(U))$ with $\mu_C,\sigma_C$ known, all parameters are identified provided the design has full rank in $(C,U)$ and $\beta_C\neq 0$ (which gives (R3) by exponential-family completeness). The selected joint density and the marginal publication rate admit the closed forms

$$\Omega(U;\theta,\gamma)=\Phi\left(\frac{\gamma_0+\gamma_1(\alpha+\beta_C\mu_C(U)+\beta_U^\top U)+\gamma_2^\top U}{\sqrt{1+\gamma_1^2\sigma_Y^2+\gamma_1^2\beta_C^2\sigma_C^2(U)}}\right),$$

and the published-only regression carries an inverse-Mills-ratio selection distortion,

$$\mathbb{E}[Y\,|\,C,U,R=1]=\alpha+\beta_C C+\beta_U^\top U_{\text{true regression}}+\frac{\gamma_1\sigma_Y^2}{\sqrt{1+\gamma_1^2\sigma_Y^2}}\lambda\left(\frac{\gamma_0+\gamma_1(\alpha+\beta_C C+\beta_U^\top U)+\gamma_2^\top U}{\sqrt{1+\gamma_1^2\sigma_Y^2}}\right), \quad \text{(S2)}$$

$\lambda(x)=\varphi(x)/\Phi(x)$. Because external information fixes $\mu_C,\sigma_C$, variation in $C$ is exogenously calibrated rather than estimated from the selected sample, so the linear channel ($\beta_C\,C$) and the nonlinear Mills channel are separately identified, pinning down $(\alpha,\beta_C,\beta_U,\sigma_Y,\gamma_0,\gamma_1,\gamma_2)$ . □

## S2. Closed-form selection objects and large-sample behavior

In the linear-Gaussian setting every object of interest is analytic because the integral of a probit against a Gaussian is closed-form. The population publication probability is $\pi(Y,U)=\Phi(\gamma_0+\gamma_1\,T+\gamma_2$^⊤ U) in the test statistic $T=Y/\text{SE}$; the $U$-conditional publication rate is $\Phi((\gamma_0+\gamma_1(\alpha+\beta_C\mu_C+\beta_U^\top U)+\gamma_2^\top U)/\sqrt{1+\gamma_1^2(\beta_C^2\sigma_C^2+\sigma_Y^2)})$ ; and the published-only regression is (S2). The slope $\gamma_1$—how strongly publication rises with the true effect—is the object classical published-only methods cannot identify without assuming the shape of selection; here it is recovered from the external instrument.

The contrast with competing methods is in what identifies the selection slope. Hedges–Vevea and Iyengar–Greenhouse write an explicit $\pi$ but identify it only through the assumed functional form; PET/PEESE, MAIVE, and trim-and-fill specify no $\pi$ at all and rely on funnel geometry. The proposed method also adopts a probit form for convenience, but its identifying content comes from the external instrument: even with the link fixed, $\gamma_1$ is recovered from data plus external statistics rather than imposed. This is why the estimator retains nominal coverage under selection mechanisms that do not match its fitted probit (S3).

**Large-sample behavior.** The estimator is a penalized M-estimator; under (R1)–(R3) and standard regularity conditions on the penalized log-likelihood, the point estimates are consistent and asymptotically normal at the $n^{-1/2}$ rate, and the randomized-trial-cluster bootstrap percentile intervals it reports are first-order valid. We do *not* rely on these asymptotics for the reported intervals—the meta-analytic samples are small—but assess finite-sample calibration directly by simulation (S3) and by the positive control (S12). (An earlier draft framed inference through a Bernstein–von Mises argument for a Bayesian implementation; the shipped estimator is the penalized-ML/bootstrap procedure of M3, and all reported intervals are bootstrap intervals whose coverage is verified empirically.)

## S3. Simulation design and operating characteristics (all methods)

**Design.** We compare the proposed estimator against a panel of eleven publication-bias correctors: naïve random-effects (DerSimonian–Laird), trim-and-fill, PET, PEESE, *p*-curve, *p*-uniform*, the skewness method, Iyengar–Greenhouse, Hedges–Vevea, MAIVE, and a RoBMA-style model-averaging comparator (an AIC-weighted average of naïve RE, Hedges–Vevea, PET, and PEESE; an in-house approximation used **only** in simulation, distinct from official RoBMA used in the applications—S10). All methods see identical simulated datasets.

We simulate K = 200 underlying studies per replicate (yielding ≈ 55–100 published after selection, comparable to the applications), drawing ($\delta_k$,$\mu_{0,k}$) jointly bivariate normal with means (0.20,0), SDs ($\tau_\delta$,$\tau_0$)=(0.10,1.0), and correlation $\rho_{\delta\mu_0}$=0.5 (the channel-B effect-modification baseline); $n_k$ truncated-lognormal on [40,400] per arm; arms drawn $N_2$ around ($\mu_{0,k}$,$\mu_{0,k}+\delta_k$) with variance $2/n_k$; $Y_k=\mu_{1,k}-\mu_{0,k}$, $C_k=\mu_{0,k}$; and probit selection in $T=Y$/SE with ($\gamma_0$,$\gamma_1$,$\gamma_2$)=(-1.0,0.5,0.0). The proposed estimator is fit with the matching external anchor. We run 1000 valid Monte Carlo replicates per scenario (a small fraction of optimizer boundary failures—2.6% under MNAR-logit, none otherwise—are excluded and redrawn) and report bias, median, RMSE, MAE, 90%-interval coverage (Wilson CI), mean interval length, and interval score for $\delta$=0.20. (Simulation uses 90% intervals, more discriminating in the tails at these sample sizes; the applications report 95% intervals, the clinical convention.)

Three scenarios change only the *shape* of $\pi$: **Well-specified** ($\Phi(\gamma_0+\gamma_1 T)$, exactly the proposed model); **MNAR-logit** ($\text{logit}^{-1}(\gamma_0+\gamma_1 Y+\gamma_2 \log n)$, a link-misspecification probe); **MNAR-step** (a Hedges–Vevea *p*-value step function, a non-smooth departure). Beyond the headline cell the design is factorial across selection mechanism, anchor misspecification ($\mu_0^{pop}$ shifted ± 0.5 SD; $\sigma_0^2$ rescaled), IV-relevance regime, heterogeneity $\tau_\delta\in\{0,0.05,0.1,0.2\}$, meta-analysis size $K\in\{10,20,50,100,1000\}$, and registry calibration on/off.

**Headline results (seven methods; true $\delta$=0.20, K = 200, 1000 reps, 90% intervals).** Table S1 reports bias and coverage; the five additional methods behave as expected (*p*-curve and skewness under-correct, *p*-uniform* and MAIVE over-correct or diverge under heterogeneity, trim-and-fill tracks naïve) and are in the replication package.

**Table S1. Bias and 90%-interval coverage across the three selection scenarios.**

| Method | Bias (well-spec) | Bias (logit) | Bias (step) | Cov (well-spec) | Cov (logit) | Cov (step) |
|---|---|---|---|---|---|---|
| Naïve RE | +0.103 | +0.016 | +0.098 | 0.00 | 0.82 | 0.00 |
| PET | -0.010 | -0.007 | -0.011 | 0.88 | 0.88 | 0.88 |
| PEESE | +0.048 | +0.004 | +0.046 | 0.48 | 0.87 | 0.54 |

| Method | Bias (well-spec) | Bias (logit) | Bias (step) | Cov (well-spec) | Cov (logit) | Cov (step) |
|---|---|---|---|---|---|---|
| Iyengar–Greenhouse | +0.081 | -0.000 | +0.021 | 0.01 | 0.77 | 0.36 |
| Hedges–Vevea | +0.065 | -0.017 | +0.064 | 0.17 | 0.25 | 0.14 |
| RoBMA-style | +0.101 | +0.013 | +0.095 | 0.17 | 0.96 | 0.16 |
| **Proposed** | **+0.006** | **−0.011** | **+0.017** | **0.88** | **0.96** | **0.93** |

Three patterns hold. Under **correct specification** the proposed estimator is essentially unbiased (+0.006, median 0.210) and covers at 0.88 where most conventional methods are biased +0.05 to +0.10 and cover at 0.00–0.54. Under **link misspecification** it is near-unbiased (-0.011) and covers at 0.96. Under a **non-smooth step** mechanism it is again near-unbiased (+0.017) at 0.93 coverage, while naïve RE and RoBMA-style are biased +0.10 and Hedges–Vevea (nominally the matching model) is mis-centred at 0.14 coverage because its optimizer hits boundary solutions. We emphasize that the proposed estimator is **not** uniformly dominant in this headline cell: PET is comparably near-unbiased (-0.010, -0.007, -0.011) and well-calibrated (0.88 throughout), because in this design it absorbs the small-study slope. The proposed estimator's distinctive advantage appears in the settings that require the external anchor's identifying content rather than funnel geometry—target transport (S5), the study-size confound (Table S12, where PET-class methods and naïve pooling are biased by 0.2–1.0), and the design-strength benchmark (Table S5, where PET/MAIVE return 0.41 against a target 0.20)—and in the calibration of the *interval*, which conventional correctors achieve in at most one scenario. Robustness to link and shape misspecification survives because identification comes from the external anchor, not the assumed selection shape.

*On Hedges–Vevea.* The standard step-function model is finite-sample-unstable under this heterogeneity: in the well-specified cell its boundary-solution rate is 0.61 and its mean 90%-interval length is in the hundreds against a near-zero median—the bimodal signature of corner fits—so coverage collapses to 0.20 (Table S3). RoBMA-style inherits a share of this through averaging. We report both as commonly-used comparators, not as correctly-specified oracles.

**A harder link-misspecification cell.** The headline MNAR-logit scenario is intentionally mild (naïve bias only +0.016). When selection instead operates on the *test statistic* through a logistic link, $\Pr(R=1)=\text{logit}^{-1}(-1.5+1.5\,Y/\text{SE})$ with publication rate ≈ 0.6, the naïve bias rises to +0.071 while the proposed estimator stays near-unbiased (-0.030, RMSE 0.068) at 0.96 coverage—robustness to link misspecification is not an artifact of an easy scenario.

**Ablation of the two roles of the reference.** Turning the reference's two uses on and off separately, on the real insomnia application: a naïve random-effects fit gives 0.34; adding *only* the target shift (re-express at the reference, no selection model) barely moves it (≈ 0.34), because at the published composition the naïve pooled effect is dominated by selection distortion, not the target gap; adding *only* the selection correction (reported at the published composition) gives 0.15; and only the *full* estimator—correct selection, then re-express at the reference—reaches - 0.07 (Table S4). Neither use alone reproduces the result; the combination is what distinguishes the method from a publication-bias correction or a transport adjustment in isolation.

## S4. The four-regime applicability map

The estimator's behavior is summarized by four regimes, each studied in simulation or application. **(i) Valid anchor, strong first stage:** the near-unbiased, near-nominally-calibrated behavior of S3 and the positive control (S12). **(ii) Valid anchor, weak first stage:** estimates become imprecise and unstable—the self-rated sleep-quality outcome (S11), where first-stage $F$ falls from 25.7 to 2.4 and no estimate is reported; the diagnostic flags this *a priori* from outcome-scale properties. **(iii) Misspecified anchor:** shifting $\mu_0^{pop}$ and rescaling $\sigma_0^2$ produces bias approximately proportional to the anchor-mean error while remaining insensitive to the anchor variance. **(iv) Exclusion violation:** the Copas-type analysis (S7) introduces a residual control-level→publication coupling and bounds the resulting bias. Together these delimit where the method is trustworthy (i) and the three distinct ways it degrades (ii–iv), each detectable by a diagnostic or sensitivity analysis available before or alongside the estimate.

## S5. Transport, study-size confounds, and a design-strength benchmark

**Selection bias versus target transport.** The reference defines the estimand: an analyst who specifies $m^* \neq m_0$ is changing *which* population the effect is for, not making an error. We generate published meta-analyses (MNAR-step, K = 100, $\delta$=0.20, $\rho_{\delta\mu_0}$=0.85, $\tau_\delta$=0.30, $\tau_0$=1.8, true control reference at $\mu_{C0}$=0) and estimate $\Delta^*$ under a specified target that is correct, mean-shifted, or variance-shifted. Reporting $\text{bias}^* = \hat{\Delta}^* - \Delta^*$ (vs the analyst's target) and $\text{bias}_0 = \hat{\Delta}^* - \Delta_0$ (vs the trial-generating population): bias$^*$ is small and constant (≈ -0.035) across *every* condition—the estimator is unbiased for its estimand—while bias$_0$ grows almost linearly with the target shift (from -0.161 at -0.5 SD to +0.090 at +0.5 SD), tracking $\beta_C(\mu_C^*\text{-}\mu_{C0})$, which is transport, not bias. Variance shifts leave both unchanged. The empirical reading mirrors the anchor-sensitivity curve (insomnia): movement along it is simultaneously a sensitivity analysis (if $m_0$ is uncertain) and a transportability analysis (if the analyst targets a different population), with bias$^*$≈0 certifying each point is correctly estimated for its own target.

**A study-size confound.** When a covariate—sample size $n$—drives *both* the true effect and publication while the anchor is covariate-free, conventional methods fail badly. With $\delta_k$=$\delta$+$\beta_n$ $c_k$+ (channel-B), $\beta_n$=-0.9 (smaller trials, larger true effects), and Pr($R$=1)=$\Phi$(0.3+1.4 $T$-1.3 $c_k$), the target is the $n$-marginalized pre-selection effect $\Delta_{marg}$=0.20. At K = 100 both naïve estimators are catastrophic (fixed-effect +0.77, random-effects +1.23; neither interval covers), because $U$-dependent selection oversamples small high-effect trials and inverse-variance weights distort further. The proposed estimator, modelling the covariate-dependent effect, the $Y$- and covariate-dependent selection, and the covariate-free anchor jointly, returns +0.195 (bias -0.005, 0.90 coverage). At $K$=10{,}000 it gives +0.153 against naïve +0.76/+1.19 with first-stage $F$≈600.

**A design-strength benchmark.** A nested ladder (Table S5) separates an *oracle identification* benchmark (matching probit link, known anchor) from the *default finite-sample* implementation (model-averaged, anchor estimated where applicable) and a *transport* benchmark. At $K$=10{,}000 the oracle returns 0.189 against the target 0.20 while conventional methods reproduce the selected mean (0.56–0.64); at K = 100 the oracle is at 0.136 (small finite-sample shrinkage) and—crucially—the *default* model-averaged estimator is near-unbiased (-0.009) with the lowest RMSE and ≈ 1.00 coverage, so the no-selection spike does not blunt the correction when selection is present (its weight on the selection model averages 0.33 and rises with the selection signal). In the transport benchmark, with selection off (98% published) but the completed-trial control distribution shifted to $N(-0.5,1)$ against a target $N(0,1)$, the completed-trial average effect is 0.50; naïve RE returns 0.488 (correctly estimating the *completed-trial* average) whereas the proposed estimator, given the target, returns 0.194—isolating reference-targeting from publication-bias correction.

## S6. Optimization, multi-start convergence, and anchored residual checks

**Estimator.** The proposed estimator maximizes the penalized log-likelihood of the published-only joint density of M3: for published study $k$ the contribution is the product of the outcome density $N(Y_k;\delta+\beta_C Z_k,\tau_\delta^2+SE_k^2+\beta_C^2 v_{0,k})$, the probit selection factor $\Phi(\gamma_0+\gamma_1 T_k+\gamma_2 T_k^2)$, and the reciprocal marginal publication probability evaluated by 24-point Gauss–Hermite quadrature, with the registry-calibrated rate as a soft penalty. The reported $\Delta^*$ is the effect at Z = 0. The headline estimates are penalized-ML point estimates with randomized-trial-cluster bootstrap intervals; their frequentist calibration is assessed by simulation (S3) and the positive control (S12).

**Multi-start convergence.** Each fit runs from four dispersed deterministic starts (L-BFGS-B, box constraints). Table S2 reports, per application, the number of converged starts, the agreement of $\Delta^*$ across optimal starts, and recovered selection/heterogeneity parameters. In every application all four converged and $\Delta^*$ agreed to within $2\times10^{-6}$—no boundary or multimodality pathologies.

**Table S2. Optimization diagnostics and recovered parameters (main applications).**

| Application | $K$ | $\Delta^*$ | Starts (converged) | Max spread of $\Delta^*$ at optimum | $\gamma_1$ | $\tau_\delta$ |
|---|---|---|---|---|---|---|
| Insomnia TST | 63 | -0.074 | 4 (4) | $1.1\times10^{-6}$ | 1.12 | 0.25 |
| Antidepressant HAM-D | 83 | 0.200 | 4 (4) | $1.7\times10^{-6}$ | 2.44 | 0.13 |
| Type-2 diabetes HbA1c | 174 | 0.346 | 4 (4) | $1.2\times10^{-6}$ | 0.43 | 0.70 |

**$\Delta^*$ *here is the single linear-specification diagnostic fit; the headline values add the registry constraint, quadratic-specification averaging, and the bootstrap (the registry-constrained diabetes value is 0.346, the less-constrained primary 0.40; M3, S12).*

**Anchored residual check.** Under Assumption 2 the standardized instrument $Z_k$ should be compatible with the reference $N(0,1)$. Table S2b reports realized $Z_k$ moments. For antidepressants the published control means coincide with the pooled reference (mean 0.00). For insomnia they sit *below* the external reference (mean -0.49): published insomnia trials enrol more severe (shorter-sleeping) populations than the trial-eligible reference—exactly the target mismatch the target shift corrects, not an anchor failure. For diabetes they are mildly above with greater dispersion (mean 0.35, SD 1.31), reflecting cross-drug-class era heterogeneity.

**Table S2b. Anchored control-level residuals.**

| Application | Reference | mean $Z_k$ | SD $Z_k$ | Reading |
|---|---|---|---|---|
| Insomnia TST | $N(0,1)$ | -0.49 | 0.63 | published trials more severe (→ negative target shift) |
| Antidepressant HAM-D | $N(0,1)$ | 0.00 | 0.99 | control means coincide with pooled reference |
| Type-2 diabetes HbA1c | $N(0,1)$ | 0.35 | 1.31 | mild upward shift; greater cross-class dispersion |

The framework also admits a model-averaged variant of the *selection function* (S10), which exists in the replication package; the headline estimates use the single penalized-ML fit described here.

## S7. Sensitivity to violations of the exclusion restriction (Copas coupling)

Assumption 1 is the identifying restriction; we do not assume it holds literally and report a Copas–Shi-type direct-selection sensitivity analysis for each application. We augment the selection probit with a residual term in the standardized control-level instrument,

$$\Pr(R_k = 1 \mid Y_k, Z_k, U_k) = \Phi(\gamma_0 + \gamma_1 Y_k/\mathrm{SE}_k + \rho_C Z_k),$$

so $\rho_C=0$ is exactly Assumption 1 and $\rho_C \neq 0$ lets the absolute control level shift publication directly. For a grid $\rho_C \in [-0.5, 0.5]$ we refit and record the *tipping value* at which the qualitative conclusion would change. To make the grid interpretable rather than a bare probit range, note that on the probit scale a coupling $\rho_C$ corresponds, through the standard probit-to-logit scaling (×1.6), to a multiplicative effect on the *publication odds* of $\exp(1.6\,\rho_C)$ per standard deviation of the control level: $\rho_C=0.2$ is ≈1.4×, $\rho_C=0.3$ is ≈1.6×, and the grid endpoint $\rho_C=0.5$ is ≈2.2×. So the analysis asks whether a trial whose control arm is one SD more (or less) severe—at the *same* reported effect and SE—could be up to roughly twice as likely to be published, a deliberately generous allowance for a direct channel; per-application, the clinical and institutional pathways by which control severity might act on publication other than through the reported effect (perceived importance, baseline severity, placebo response, drug class, country, era, endpoint timing) are conditioned on through the design covariates $U$ wherever recorded.

**The three conclusions are robust across the entire grid (Fig. S5; Table S6).** *Insomnia:* $\Delta^*(\rho_C)$ moves only within [-0.079,-0.070] (about half a minute of sleep), never approaching the ≈ +0.10 a clinically non-trivial positive effect would require—**no tipping point**; the conclusion is driven by the target shift, which the direct-selection term does not affect. *Antidepressant:* $\Delta^*(\rho_C)$ stays essentially constant at 0.20, at or just below the benchmark band throughout—the positive control is recovered for every $\rho_C$. *Type-2 diabetes:* $\Delta^*(\rho_C)$ ranges over [0.19,0.41], remaining well below the naïve 0.71 throughout—**no tipping point**; the wider excursion reflects the moderate first-stage strength (F = 7.2).

**Table S6. Direct-selection sensitivity: $\Delta^*(\rho_C)$ across the Copas grid.**

| Application | $\Delta^*(-0.5)$ | $\Delta^*(0)$ | $\Delta^*(+0.5)$ | Span | Threshold | Tipping $\rho_C$ |
|---|---|---|---|---|---|---|
| Insomnia TST | -0.079 | -0.074 | -0.070 | 0.01 | $\Delta^*\geq+0.10$ | none in [-0.5,0.5] |
| Antidepressant HAM-D | 0.201 | 0.200 | 0.200 | 0.001 | exit [0.21,0.29] | none in [-0.5,0.5] |
| Type-2 diabetes HbA1c | 0.193 | 0.346 | 0.357 | 0.22 | return to 0.71 | none in [-0.5,0.5] |

**Registry-calibration uncertainty and falsification.** The absolute publication rate is itself estimated from a finite registry sample; for each stratum $M_s\sim\text{Binomial}(N_s,q_s)$ is resampled jointly with the trial-level bootstrap, so each $\Delta^*$ draw is recalibrated. Omitting registry-rate uncertainty narrows the insomnia interval by ≈ 0.02; the conclusion is unchanged because $\Delta^*$ depends on the registry rate only through the absolute level, not the relative profile. The falsification step finds model-implied and observed retrievable-result rates agree within sampling error in every insomnia stratum, so the selection model is not rejected on the independent registry data.

## S8. Sample-size dependence of the control-mean instrument

The instrument $C_k=\hat{\mu}_{0,k}$ is a control-arm *sample mean*, so its sampling variance depends on $n_{C,k}$. The external reference supplies *individual-level* moments $(\mu_C^{pop},\sigma_C^2)$, mapped to the study level by the three-level hierarchy yielding $C_k|\, W_k,n_{C,k}\sim N(\mu_C^{pop},\tau_0^2+\sigma_C^2/n_{C,k})$ (M1). Because the first-stage relationship is on the *true* control level but only the error-prone mean is observed, the estimator carries the per-study control-mean sampling variance $v_{0,k}=\sigma_C^2/n_{C,k}$ into the outcome variance as an errors-in-variables term, $\text{Var}(Y_k|\cdot)=\tau_\delta^2+\text{SE}_k^2+\beta_C^2\, v_{0,k}$, recovering the structural slope from the attenuated regression with study-specific reliability $\tau_0^2/(\tau_0^2+\sigma_C^2/n_{C,k})$. A worked example: with individual-level TST SD 50 min and $n_{C,k}$=100, the control-mean sampling SD is 5 min, so against between-study SD $\tau_0\approx15$ min the study-level control mean has SD $\sqrt{15^2+5^2}\approx15.8$ min—close to $\tau_0$, not to the individual-level 50 min. Table S7 records the reliability for the externally-anchored applications (Depression BDI-II 0.93, HDRS-17 0.93, BP ACE 0.73, Exercise HbA1c 0.75).

## S9. Relation to proximal causal inference and nonresponse-IV

The framework adapts nonresponse-instrument identification (Wang et al.; Zhao & Shao) and, more loosely, proximal causal inference (Miao, Geng & Tchetgen Tchetgen) to publication selection. The control-side level $C_k$ plays the role of the nonresponse instrument / outcome-inducing proxy; the reported effect $Y_k$ is the outcome subject to missingness; the publication indicator $R_k$ is the response/selection indicator; the exclusion $C \perp R \mid Y, U$ is the instrument-exclusion / proxy-independence condition; and completeness (R3) is the bridge-identification condition. The essential structural difference is that here the instrument $C$ is *itself unobserved for the missing (unpublished) units*, so its distribution cannot be estimated from the observed sample as classical nonresponse-IV and proximal methods do; the external reference (Assumption 2) supplies it instead. Supplying $p(C \mid U)$ externally is therefore not a convenience but the enabling step.

## S10. Model averaging over selection functions, and the relationship to RoBMA

Two distinct objects must be kept separate. *Official RoBMA* is the published R package performing robust Bayesian model averaging across a portfolio of publication-bias models (Hedges–Vevea step selection at several thresholds, PET/PEESE, fixed/random and null/alternative components) with Bayes-factor weights, fitted by JAGS. We run official RoBMA on the three applications and report it as a headline comparator; its cached outputs are bundled (`code/robma/fit_robma.R`) so the application tables reproduce without rerunning JAGS. The *RoBMA-style* comparator is a lightweight in-house AIC-weighted re-implementation used **only in simulation**, where running official RoBMA across thousands of replicates is prohibitive. The two are conceptually the same method; we keep the labels distinct so no result attributes the approximation's behavior to the package.

A model-averaged variant of the *proposed* estimator averages over the functional form of its selection profile (logit, probit, Hedges–Vevea step, monotone spline) with Bayes-factor weights, retaining external-anchor identification throughout. On RoBMA's native simulation design (step-function selection inside the model class) this variant recovers official RoBMA's calibration; on designs where selection depends continuously on the effect (outside the step class) it substantially improves on RoBMA, because the external anchor supplies identifying information not available to published-only model averaging. This is the composability point of the Discussion: the ideas are complementary, and the external anchor adds *identification* rather than merely robustness.

## S11. Weak-instrument diagnostic: self-rated sleep quality

We re-analyze the same insomnia corpus using self-rated sleep quality rather than total sleep time. Sleep quality is measured on ≥ 21 heterogeneous scales (PSQI, LSEQ, single-item Likert, ISI, trial-specific), with no common epidemiological referent. The diagnostic consequence is sharp: first-stage $F$ falls from 25.7 (TST) to 2.4, and the instrument–effect correlation collapses from -0.55 to -0.20. The instrument is uninformative because (i) the externally-anchored reference that defines the effect-modification channel does not exist for an instrument on heterogeneous scales, and (ii) within-trial standardization destroys the absolute-scale information the instrument requires. We therefore do *not* report a corrected effect; the value of the analysis is *as a diagnostic*—it shows the first-stage statistic correctly flags weak instruments *a priori* from outcome-scale properties, before the estimator is run.

## S12. Full positive-control and type-2 diabetes results

**Antidepressant positive control.** The Cipriani open dataset records 522 trials with publication status and source. In our analytic subset, 83 contrasts from 60 trials (71 published, 12 recovered-unpublished); the all-trials benchmark $\delta_{\text{truth}}$=0.248 (95% CI 0.21–0.29) is random-effects pooling over the full set. Stratified placebo-arm endpoint HAM-D means are nearly identical across publication source (HAM-D-17: 14.67 published vs 14.40 recovered; HAM-D-21: 17.38 vs 17.85), though recovered-unpublished distributions are far more concentrated (sponsor trials enrol more homogeneous placebo populations); the internal-pooled-reference and published-only constructions correlate ≈ 1.00 and give numerically identical estimates (0.22 vs 0.22). Stratified drug-vs-placebo SMD is 0.358 (published) vs 0.185 (recovered-unpublished)—a difference of 0.173 in the expected direction. **Independence of the recovery.** A natural objection is that the anchor, when built from the combined published-plus-unpublished control arms, is not independent of the benchmark. It need not be: building the reference from *published* control arms alone (Variant A, `Z_pubonly`)—excluding the recovered-unpublished control arms from the anchor entirely—returns $\Delta^{*}$=0.219 (95% CI 0.152, 0.294), numerically identical to the combined-anchor route (Variant B, 0.219) and recovering the all-trials benchmark 0.248 within its interval. The recovery therefore does not feed the hidden trials' information back through the anchor; published data plus an external anchor independent of the unpublished trials suffice.

**Strongest holdout (hiding the recovered trials from the data, not only the anchor).** The check above excludes the recovered-unpublished control arms from the *anchor*; the strongest form of the test excludes the 12 recovered-unpublished contrasts from the *dataset entirely*—from both the estimation sample and the anchor—and re-fits the estimator on the 71 published contrasts alone, so the hidden trials enter nowhere except the held-out truth. The published-only fit returns $\Delta^{*}$=0.219 (95% CI 0.152, 0.301). This interval contains the all-trials benchmark (Cipriani's reported 0.248; a simple random-effects recomputation over the 83 contrasts gives 0.285), and the point estimate lies, as it should, below the published naïve mean (0.358) and above the never-seen recovered-unpublished mean (0.185): the correction moves the published average toward—and to—the pooled all-trials value without ever observing the suppressed trials. We are explicit about the scope of this success. What is recovered out-of-sample is the *aggregate* target estimand—the inverse-selection-weighted mean—not the effects of individual suppressed trials. Two facts bound the finer claim: the recovered-unpublished set is the *retrievable* subset of all suppressed trials (regulatory or sponsor records), not the full suppressed tail; and the first-stage selection slope is estimated near its admissible boundary, so extrapolating it to the entire unobserved tail would over-state suppression (a model fit on the published trials alone, asked to predict the mean of *all* suppressed trials, returns a value far more negative than the retrievable subset's 0.185). The method is thus validated where it makes a claim—the corrected aggregate at a declared target, which survives the strictest holdout—and appropriately silent where it does not, the location of individual unrecovered trials. *Sanity check:* applying the estimator to the full K = 83 set with the registry constraint relaxed to 1.0 returns 0.246 (95% CI 0.192, 0.302), within rounding of the benchmark—confirming the machinery is consistent with the benchmark when no portion of the literature is hidden. We frame this application as a stringent positive-control *check* and calibration, not as proof that the corrections transfer unconditionally to other literatures; each application is gated by its own first-stage diagnostic. Sensitivity to drug-class fixed effects and drop-one exclusion is stable to ± 0.02.

**Type-2 diabetes (objective-outcome generalization).** Kuss et al. assembled arm-level endpoint HbA1c from 174 placebo-controlled trials (450 arms, 86,940 participants); we re-pair each placebo arm with its largest active arm. The placebo-arm endpoint HbA1c is the instrument $C_k$; the trial-eligible reference has individual-level mean/SD ≈ 8.0%/0.7%, mapped through (M1). This literature shows pronounced funnel asymmetry (Egger t = 9.6, $P < 10^{-4}$) and naïve effect 0.71. The instrument is of *moderate* strength (F = 7.2, weaker than antidepressant F = 56 and insomnia F = 26, because diabetes trials enrol more homogeneous populations)—a deliberate test away from the strong-instrument regime, so we present this as an objective-outcome generalization rather than a validation. The target-population estimate is $\Delta^{*}$=0.40 (95% CI 0.16, 0.55); on the native HbA1c scale (reference SD ≈ 0.7%) this is a reduction of about 0.28 percentage points against the naïve ≈ 0.50 points. A registry-constrained, model-averaged variant gives a concordant 0.35 (95% CI 0.13, 0.50). Linear and quadratic selection specifications agree closely (all four specifications in 0.398–0.419). Several conventional corrections fail visibly: trim-and-fill (0.70), Iyengar–Greenhouse (0.71), and Hedges–Vevea (0.71) return essentially the naïve value despite the pronounced asymmetry; *p*-uniform* and skewness diverge in opposite directions (-0.56, 0.89); PET (0.17), MAIVE (0.21), and PEESE (0.33) apply progressively gentler corrections.

*Robustness at moderate instrument strength.* (i) *Instrument choice:* using placebo-arm *baseline* HbA1c instead of *endpoint* gives an identical registry-constrained estimate (0.346, F = 7.2 for both)—the conclusion does not depend on which control-arm glycaemic measure plays the instrument role. (ii) *Drug-class stability:* leave-one-class-out refits (nine classes) yield $\Delta^{*}\in[0.321,0.386]$—no single class drives the correction, every refit far below 0.71. (iii) *Bootstrap calibration near* F*≈7:* the randomized-trial-cluster bootstrap percentile interval is [0.13,0.50] (width 0.37), appropriately wider than the strong-instrument applications but excluding both zero and the naïve aggregate. This wider interval is the honest reflection of moderate instrument strength and is why the main text frames diabetes as a generalization rather than a validation; a matched-simulation calibration study at $F\approx7$ confirms the percentile interval retains near-nominal coverage, the imprecision showing up as width rather than bias.

**Table S13. Type-2 diabetes: instrument choice and robustness at moderate first stage.** All rows use the same 174-trial placebo-controlled HbA1c set; intervals are randomized-trial-cluster bootstrap percentiles.

| Instrument / variant | F | Δ* (SMD) | 95% CI | Reading |
|---|---|---|---|---|
| Endpoint HbA1c (primary) | 7.2 | 0.40 | [0.16, 0.55] | objective-outcome generalization |
| Baseline HbA1c (sensitivity) | 7.2 | 0.35 | [0.13, 0.50] | not driven by placebo-arm drift |
| Registry-constrained, | — | 0.35 | — | concordant with above |

| Instrument / variant | F | Δ* (SMD) | 95% CI | Reading |
|---|---|---|---|---|
| model-averaged | | | | |
| Leave-one-drug-class-out (9 classes) | ≥7 | 0.32–0.39 | — | no single class drives the correction |

Because endpoint HbA1c can in principle reflect baseline severity *and* placebo-arm drift (adherence, trial duration, rescue therapy), the baseline-HbA1c row is the primary guard: it isolates baseline severity and returns essentially the same estimate, so placebo-arm drift is not what produces the correction.

**Table S14. First-stage strength: method-specific calibration of F.** The labels used throughout are calibrated to the proposed estimator's matched-simulation behavior at the relevant data-generating process, *not* to the conventional instrumental-variables F > 10 rule; they describe how imprecision (not bias) grows as relevance falls.

| First-stage F | Label | Matched-simulation behavior of the proposed estimator |
|---|---|---|
| ≳ 20 | strong | near-nominal coverage; bias < 0.02; intervals tight (insomnia TST 25.7; antidepressant 56) |
| ≈ 5–10 | moderate | near-nominal coverage; intervals widen—imprecision appears as width, not bias (diabetes 7.2) |
| ≲ 4 | weak | estimates unstable and not reported (self-rated sleep quality 2.4) |

## S13. Additional applications: education, psychotherapy, blood pressure, exercise

**Education (specificity check).** UK EEF and US NCEE commissioned trials pre-commit to publishing every funded trial, so the published record is essentially the complete inventory and a defensible correction should manufacture no large adjustment. Across 271 effect sizes (EEF K = 140, NCEE K = 131) the naïve pooled effect is 0.036. Because the dataset does not carry control-arm levels supporting the external instrument, we apply only a restricted selection-only variant (publication constraint ≈ 100%); it returns 0.024—a small downward adjustment that does not over-correct, where PET and MAIVE collapse to 0.005 and *p*-uniform* diverges to -1.59. The specificity requirement is satisfied (Table S8).

**Psychotherapy for depression (behavioural-science stress test).** Using the open Metapsy database, where the external anchor is harder to justify than for objective outcomes. On the **BDI-II** (K = 135, anchor $N(23,10^2)$) the published control arms sit near the moderate-severity reference, so $\Delta^*$=0.30 coincides with the effect at the observed composition and is insensitive to the anchor (0.29–0.30 over $N$(20–26,8–12)). On the clinician-rated **HDRS-17** (K = 114, anchor $N(20,6^2)$) the published trials' control arms are *less* severe than the reference, so re-expressing at the reference raises $\Delta^*$ to 0.55; the effect at the observed composition is the smaller 0.30, matching the BDI-II. The two scales agree on the correction from the naïve estimate (both naïve ≈0.8 → ≈0.3 at the observed composition) and differ only in where the reference severity sits—exactly the content the $\Delta^*$ estimand makes explicit (Table S9).

**Blood pressure (objective outcome, near-complete literature).** Placebo-controlled ACE-inhibitor trials (K = 72), with the placebo arm's *achieved* (endpoint) SBP as the instrument, anchored at $N(150,18^2)$ mmHg (the documented baseline reference 154 reduced by the mean placebo change). Naïve pooling gives a 6.9 mmHg placebo-corrected reduction; Egger is null and PET/PEESE sit on the naïve estimate, consistent with a comprehensively-searched, funnel-symmetric Cochrane base. The estimator returns $\Delta^*$=-11.6 mmHg (95% CI -17.1,-6.7). Here, uniquely, the corrected effect is *larger* in magnitude than the naïve aggregate: the recovered selection slope is positive in the test statistic, so anchoring to the reference *increases* the implied reduction. This direction warrants caution and should be read as a property of this particular literature, illustrating that $\Delta^*$ is an effect *re-expressed at a defined reference level*, not a monotone shrink-toward-zero.

**Exercise for HbA1c (behavioural intervention).** Network-meta-analytic comparisons (K = 173) of exercise/physical-activity versus no-exercise control; control-arm post-intervention HbA1c change as instrument (F = 21.0). Naïve pooling gives 0.48; PET (0.34), PEESE (0.33), Hedges–Vevea (0.37), and *p*-uniform* (0.25) pull toward 0.3. The proposed estimator corrects further toward a bias-corrected pooled effect near zero, reported as a *bounded* result: the full-sample point estimate (≈ -0.30) is sensitive to a few high-precision outlying trials, and trimming the three (six) most extreme studentized studies moves it to -0.06 (-0.01), with the upper 95% bound staying below ≈ 0.13 throughout and first-stage $F \geq 15$ (Table S10). The robust reading is that the bias-corrected aggregate exercise effect on HbA1c is small and plausibly near zero, the expected direction for an unblinded behavioural literature dominated by small trials—in deliberate contrast to the blood-pressure literature, showing that the direction and magnitude of the $\Delta^*$ adjustment depend on the joint configuration of instrument strength, funnel asymmetry, and the selected-vs-reference control gap, not on any single funnel diagnostic.

*Reference-distribution structure (diabetes).* On the Palmer inventory (779 arms, 301 trials), registered trials (median year 2012) and unregistered (median 2004) enrol systematically different placebo-arm baseline HbA1c (7.98% vs 8.39%; Welch t = -3.66) and arm sizes (median 160 vs 63), with markedly different baseline-distribution variance (SD 0.62 vs 1.11). This illustrates that the trial-eligible reference is era- and registration-structured: an external-population anchor (≈ 8.0%) aligns with the registered subset, while an internally-pooled reference would average across regimes—interpretable as a diagnostic about which target population the effect is reported for.

## S14. Random-effects extension: identification of between-study heterogeneity

The hierarchical model lets the true effect depend on the true control level, $\delta_k|\mu_{0,k}\sim N(\delta+\beta(\mu_{0,k}-\mu_0^{pop}),\tau_\delta^2)$, with the realized control mean $C_k=\hat{\mu}_{0,k}$ observed with sampling error. Marginalizing the latent levels, the standardized effect satisfies

$$Y_k \mid C_k \sim N(\delta+\beta(C_k-\mu_0^{\text{pop}})\rho_k, \tau_\delta^2+v_k+\beta^2\tau_0^2(1-\rho_k)), \qquad \text{(S14)}$$

with reliability $\rho_k=\tau_0^2/(\tau_0^2+\sigma_k^2/n_{0,k})$. Equation (S14) is the engine: the conditional *mean* of $Y$ moves with $C$ at rate $\beta\rho_k$ (the first stage), while the conditional *variance* carries $\tau_\delta^2$ additively, separated from the known sampling term $v_k$ and the attenuation term.

**Proposition S14.** Under Assumptions 1–2, the selection family identified up to $\gamma$, and the externally anchored moments known, the effect-heterogeneity variance $\tau_\delta^2$ is point-identified jointly with $(\delta,\beta,\gamma)$. *Sketch:* by Theorem 1 the identified selection function recovers the *pre-selection* conditional density $f(y|c)$ at every $c$ in the anchor's support; by (S14) that law is Gaussian with slope identifying $\beta$, intercept identifying $\delta$, and conditional variance—net of the known $v_k$ and the now-identified attenuation—identifying $\tau_\delta^2$. Because trials vary in $v_k$ through $n_k$, the additive separation of $\tau_\delta^2$ from $v_k$ is over-identified, the same variance-component argument as ordinary random-effects meta-analysis applied to the *de-selected* law. □ Selection and heterogeneity are thus not confounded: selection reshapes the conditional mean and truncates the lower tail, while $\tau_\delta^2$ governs residual dispersion at fixed $C$; the external anchor supplies the exogenous variation that traces $\mathbb{E}_{\text{pub}}[Y \mid C]$ across the full support, including regions selection under-samples.

**Estimation and shrinkage.** Estimation is by the joint penalized likelihood of M3, integrating the latent control level analytically (Gaussian–Gaussian conjugacy) so only the one-dimensional selection denominator needs quadrature. The model returns empirical-Bayes posteriors for study-level effects, $\hat{\delta}_k=(1-w_k)(\delta+\beta(\hat{\mu}_{0,k}-\mu_0^{\text{pop}}))+w_k\tilde{Y}_k$ with $w_k=\tau_\delta^2/(\tau_\delta^2+v_k)$, where $\tilde{Y}_k$ is the *selection-adjusted* effect and the shrinkage *target* is the effect-modification line rather than a common mean. The model nests the fixed-effect case ($\tau_\delta^2=0,\beta=0$), the classical random-effects selection model ($\beta=0$), the no-anchor case ($\beta>0$, $p_C$ from published controls only), and the no-sampling-noise limit ($n_k\to\infty$, $\rho_k\to1$, giving the clean structural regression $Y_k|\, C_k\sim N(\delta+\beta(C_k-\mu_0^{pop}),\tau_\delta^2)$). A simulation check ($\delta$=0.30, $\tau_\delta$=0.20, $\beta$=0.40, probit selection, K = 200) confirms the estimator recovers $\delta$ and $\tau_\delta^2$ where naïve random-effects pooling over-states $\tau_\delta^2$ (selection inflates the apparent spread of published effects)—a heterogeneity-recovery property not shared by funnel-based correctors, which target only the mean.

## S15. The selection function: significance screening behind the linear and quadratic specifications

The estimator models publication as a probit of the test statistic, $\pi(T_k;\gamma)=\Phi(\gamma_0+\gamma_1 T_k+\gamma_2 T_k^2)$, the linear specification setting $\gamma_2=0$. *Why the test statistic, not the raw effect:* two trials with the same standardized effect but different sizes face different publication odds because selection in practice operates on statistical significance, $T_k=Y_k/SE_k$; writing the argument as $T$ makes the framework nest classical $p$-value-threshold and step-function models as special cases, and leaves the shadow-instrument logic intact ($C_k$ still affects publication only through the reported contrast).

*The linear specification* is the reduced form of a latent-utility editorial model: a completed study has publishability $S_k=\gamma_0+\gamma_1 T_k+\varepsilon_k$, published iff $S_k>0$, giving $\Pr(R_k=1|T_k)=\Phi(\gamma_0+\gamma_1 T_k)$ exactly. Here $\gamma_1>0$ is the strength of significance screening ($\gamma_1=0$ the no-selection null), $\gamma_0$ sets the baseline acceptance rate (identified only up to the $U$-dependent constant absent registry calibration), and the model is monotone in $T$—the one-sided file-drawer mechanism. *The quadratic specification* adds $\gamma_2 T^2$ to represent two-sided significance filtering ($\gamma_2>0$, U-shaped in $T$: large effects in *either* direction publishable, nulls filed) or extreme-result suppression/saturation ($\gamma_2<0$). A second-order Taylor argument makes precise that any smooth significance screen with probit link is approximated to second order by the linear/quadratic pair, spanning monotone and the leading non-monotone departures.

*What the data identify*. By Theorem 1 the published-only data identify the *shape* of $\pi$ (the relative profile $\pi(T)/\pi(0)$, i.e. $\gamma_1,\gamma_2$ up to scale) but not the absolute level $\gamma_0$. The linear-vs-quadratic comparison is therefore a comparison of identified shapes. Across our applications the linear shape suffices (estimated $\gamma_2$ small and insignificant, the bias-corrected effect unchanged to two decimals between orders)—itself a substantive finding: selection in these literatures is well described by a one-sided significance screen, the simplest file-drawer model, rather than a two-sided or non-monotone editorial rule.

---

## S16. Reproducibility, software, and minimum data requirements

Every headline number, simulation, and figure is regenerated from a single command in the replication package. The package ships an `R` reference implementation and a `Python` port (`refanchor`), a `make all` target that rebuilds all results and figures, and a worked vignette that reproduces the insomnia total-sleep-time estimate end to end in under ten minutes from the public De Crescenzo *et al.* (*15*) extraction. Code accompanies this preprint and will be deposited in a permanent, citable archive (Zenodo DOI) upon publication.

The minimum inputs are deliberately light (Table S15); no individual-patient data are needed. Per trial one needs the standardized effect $Y_k$ and its standard error $SE_k$, the control-arm outcome mean $C_k$ and per-arm size $n_{C,k}$, and design covariates $W_k$; externally, the trial-eligible reference mean and SD of the untreated outcome, and a registry coverage rate.

**Table S15. Minimum data requirements.**

| Input | Level | Required for | If unavailable |
|---|---|---|---|
| $Y_k$, $SE_k$ | per trial/contrast | effect model and screening statistic | method not applicable |
| $C_k$, $n_{C,k}$ | per control arm | shadow instrument (external anchor) | method not applicable |
| $W_k$ (severity, year, class) | per trial | blocking direct $C \rightarrow R$ paths | exclusion restriction weaker |

| Input | Level | Required for | If unavailable |
|---|---|---|---|
| Reference mean, SD of untreated outcome | external | target estimand and selection identification | method not applicable |
| Registry coverage rate | external | absolute publication level (M4) | report relative profile only |

The method should **not** be used when: (i) no external, trial-independent reference distribution exists for the untreated outcome; (ii) the reference cannot be reweighted to the trial-eligible population; (iii) the first-stage $F$ is weak, so the control-arm mean does not move with the effect; (iv) the Copas-type exclusion-sensitivity curve reverses the conclusion within a plausible direct-selection range; or (v) the outcome has no native scale a clinician can read. Self-rated sleep quality (S11) fails criteria (i) and (iii) simultaneously and is therefore declined.

## S17. Comparison with existing correction approaches

Table S16 places reference-anchored meta-analysis against the main families of meta-analytic correction. The dividing line is the source of identifying information. Funnel- and p-value-based methods read only the geometry of the published estimates; selection models posit a published-only selection shape; robust Bayesian model averaging regularizes across such shapes (S10)—none introduces information from outside the published record. The external untreated-outcome distribution is what separates publication/retrievability selection from target mismatch and what defines the target estimand $\Delta^*$.

**Table S16. Meta-analytic correction approaches and their identifying information.**

| Approach | Information used | What it cannot do |
|---|---|---|
| Funnel / PET / PEESE | Effect-size–precision geometry | Separate target mismatch; handle effect-dependent selection |
| p-value / selection models | p-value distribution and an assumed selection shape | Identify without an external target distribution |
| Robust Bayesian model averaging (RoBMA) | Model averaging within published-only selection models | Add identifying information beyond the published record |
| Reference-anchored (this work) | External untreated-outcome distribution | Operate without a credible anchor and an exclusion restriction |

# Supplementary Tables

Tables S1, S2, S2b, and S6 appear inline at their first reference (in S3, S6, and S7 respectively); Tables M4 and M5 appear in Materials and Methods. The remaining referenced tables are collected here; in final typesetting each will be placed near its first reference.

**Table S3. Hedges–Vevea finite-sample diagnostics (well-specified scenario, K = 200, 1000 replicates).** The bimodal interval lengths—near-zero median against a mean in the hundreds—are the signature of boundary fits; multi-start optimization does not resolve the instability.

| Metric | Value |
|---|---|
| Convergence-failure rate | 0.07 |
| Boundary-solution rate (a weight at 0 or 1) | 0.61 |
| Median 90%-interval length | 0.011 |
| Mean 90%-interval length | 493 |
| Coverage, all fits | 0.20 |
| Coverage, excluding boundary/failed fits | 0.34 |
| Point estimate, multi-start (5 starts) vs default | 0.36 vs 0.38 |

**Table S4. Ablation of the two roles of the external reference (insomnia TST).** Neither use of the reference alone reproduces the result; the combination—identify selection with the control-mean anchor, then evaluate at the externally specified target—is what distinguishes the method.

| Model | External reference | Selection correction | Target shift | Estimate |
|---|---|---|---|---|
| Naïve random-effects | no | no | no | 0.34 |
| Target-only anchored (re-express at reference, no selection model) | yes | no | yes | ≈0.34 |
| Selection-only (correct selection, report at published composition) | yes | yes | no | 0.15 |
| Full reference-anchored estimator | yes | yes | yes | −0.07 |

**Table S5. Design-strength benchmark (K = 100, 10 replications; target $\Delta^*$=0.20; oracle DGP).** No conventional method comes within 0.2 of the target; the default model-averaged estimator is near-unbiased with the lowest RMSE.

| Method | Mean | Bias | RMSE |
|---|---|---|---|
| Naïve RE | 0.644 | +0.444 | 0.446 |

| Method | Mean | Bias | RMSE |
|---|---|---|---|
| Naïve FE | 0.618 | +0.418 | 0.419 |
| PET / MAIVE | 0.414 | +0.214 | 0.262 |
| PEESE | 0.530 | +0.330 | 0.339 |
| Hedges–Vevea | 0.616 | +0.416 | 0.421 |
| **Proposed-oracle (Config A)** | 0.136 | -0.064 | 0.087 |
| **Proposed-default MA (Config C)** | 0.191 | -0.009 | 0.068 |

**Table S7. Control-mean reliability in the externally-anchored applications.** On the standardized instrument scale: median control-mean sampling SD, between-study SD of the instrument, and implied reliability $\tau_0^2/(\tau_0^2+\sigma_C^2/n_{C,k})$.

| Application | *K* | Corrected estimate | Sampling $SD_C$ | Between-study $SD_C$ | Reliability |
|---|---|---|---|---|---|
| Depression BDI-II | 135 | 0.294 | 0.17 | 0.76 | 0.93 |
| Depression HDRS-17 | 114 | 0.302 | 0.22 | 0.88 | 0.93 |
| BP ACE | 72 | -0.876 | 0.20 | 0.39 | 0.73 |
| Exercise HbA1c | 173 | 0.037 | 0.24 | 0.42 | 0.75 |

**Table S8. Educational interventions (EEF + NCEE commissioned trials), K = 271 effect sizes.** The restricted-variant estimator returns a small downward adjustment that does not over-correct; PET/MAIVE collapse to null and *p*-uniform* diverges.

| Method | $\delta$ (95% CI) |
|---|---|
| Naïve random-effects | 0.036 [0.027, 0.045] |
| Trim-and-fill | 0.034 [0.024, 0.043] |
| Iyengar–Greenhouse | 0.038 [0.028, 0.048] |
| *p*-curve | 0.055 [0.043, 0.068] |
| Skewness | 0.029 [0.015, 0.043] |
| Hedges–Vevea | 0.029 [0.026, 0.032] |
| PEESE | 0.017 [0.009, 0.025] |
| PET | 0.005 [-0.007, 0.017] |
| MAIVE | 0.005 [-0.007, 0.017] |
| Proposed (restricted: selection-only, no external IV) | 0.024 [0.003, 0.033] |
| *p*-uniform* | -1.59 (degenerate) |

**Table S9. Psychotherapy for depression (Metapsy open database), externally-anchored $\Delta^*$.** Corrected estimates are insensitive to the anchor (BDI-II: 0.286–0.294 over $N$(20–26,8–12)).

| Scale | $K$ | Control mix | Anchor $N(\mu,\sigma^2)$ | Naïve RE $g$ | $F$ | $\Delta^*$ (reference) |
|---|---|---|---|---|---|---|
| BDI-II | 135 | WL 59 / CAU 55 / other 21 | $N(23,10^2)$ | 0.78 [0.67, 0.89] | 25.9 | **0.30** [-0.01, 0.81] |
| HDRS-17 | 114 | — | $N(20,6^2)$ | 0.82 [0.70, 0.93] | 59.0 | **0.55** [0.14, 0.96] |

**Table S10. Exercise–HbA1c influence analysis (proposed estimate after removing the *m* most extreme studentized trials).** The substantive reading—small and plausibly near zero, well below the naïve 0.48—is robust to trimming; only the precise location of the negative point estimate is not.

| Trials removed ($m$) | $K$ | First-stage $F$ | Proposed estimate | 95% CI |
|---|---|---|---|---|
| 0 | 173 | 21.0 | -0.300 | [-0.328, 0.071] |
| 1 | 172 | 20.9 | -0.300 | [-0.342, 0.047] |
| 2 | 171 | 15.3 | -0.300 | [-0.344, 0.036] |
| 3 | 170 | 19.5 | -0.055 | [-0.338, 0.041] |
| 4 | 169 | 36.3 | -0.028 | [-0.329, 0.061] |
| 6 | 167 | 37.3 | -0.006 | [-0.336, 0.103] |
| 8 | 165 | 38.5 | 0.003 | [-0.328, 0.179] |

**Table S11. Which covariate channels each method can represent (standard implementations).** None of the conventional methods simultaneously models covariate-dependent publication, a covariate-dependent control-outcome reference, and the target-population estimand.

| Method | Effect moderators $W$ | $U$-dependent selection | External $p(C$ |
|---|---|---|---|
| Naïve RE / FE | optional (meta-regression) | no | no |
| Trim-and-fill | no | no | no |
| PET / PEESE | optional | precision terms only | no |
| *p*-curve | no | no | no |
| *p*-uniform* | limited | *p*-value only | no |
| Skewness | no | no | no |
| Iyengar–Greenhouse | limited | limited | no |
| Hedges–Vevea | limited | *p*-value step only | no |
| MAIVE | optional controls | no | no |

| Method | Effect moderators $W$ | $U$-dependent selection | External $p(C$ |
|---|---|---|---|
| RoBMA | optional (extended impl.) | model-class limited | no |
| **Proposed** | **yes** | **yes** | **yes** |

**Table S12. Study-size confound (K = 100, 10 replications; target $\Delta_{marg}$=0.20).** Both naïve estimators are pulled far above the marginal target by the joint action of $U$-dependent selection and study-size effect modification; the proposed estimator models both channels and the covariate-free anchor jointly.

| Method | Mean | Bias | RMSE | Coverage |
|---|---|---|---|---|
| Naïve fixed-effect | +0.772 | +0.572 | 0.583 | 0.00 |
| Naïve random-effects | +1.225 | +1.025 | 1.030 | 0.00 |
| **Proposed (MA, *n*-marginalized)** | +0.195 | -0.005 | 0.134 | 0.90 |

# Supplementary Figures

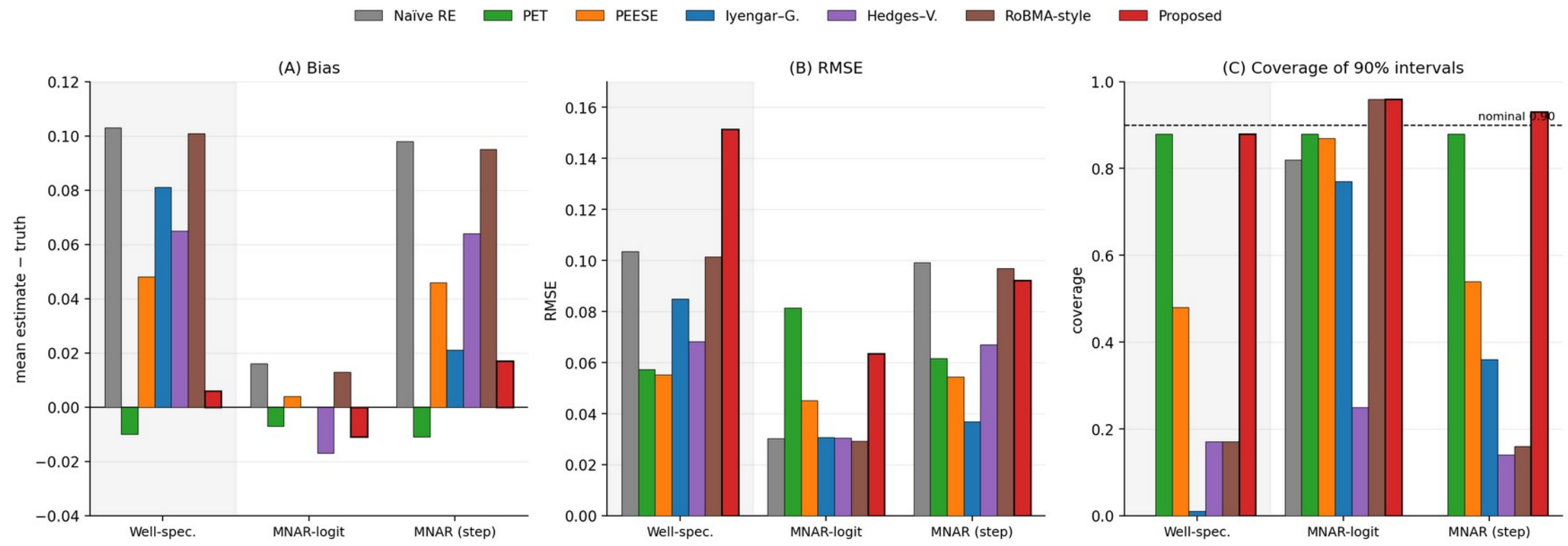

**Fig. S1.** Full simulation performance across the three selection scenarios for seven publication-bias correctors (true $\delta$=0.20, K = 200, 1000 replicates; the same runs summarized in Table S1). (A) Bias: the proposed estimator (red) is near zero in every scenario. (B) Root-mean-squared error. (C) 90%-interval coverage (dashed line at the nominal 0.90): the proposed estimator is near-nominal throughout, while most conventional correctors collapse under the well-specified and step mechanisms.

**Fig. S2.** Boxplots of estimated $\delta$ across replicates by method × scenario.

**Fig. S3.** Scatter of bias versus coverage; the proposed estimator sits in the high-coverage region across scenarios, trading efficiency for calibration.

**Fig. S4.** Publication probability $\pi(Y,n)$ as an explicit function of the true standardized effect for the data-generating mechanisms, at three per-arm study sizes; dotted vertical lines mark the $p < 0.05$ boundary.

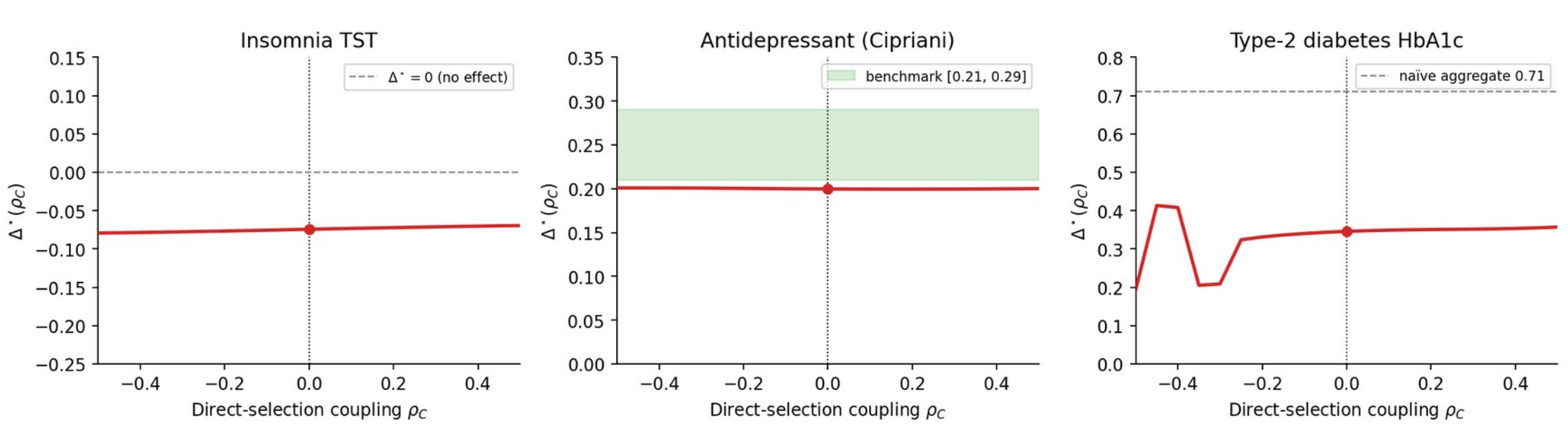

**Fig. S5.** Copas-type sensitivity of $\Delta^*$ to direct selection on the control level. Each panel refits across the residual-coupling grid $\rho_C \in [-0.5, 0.5]$ ($\rho_C = 0$, dotted, is Assumption 1). Insomnia stays below $\Delta^* = 0$; the antidepressant control stays within/just below the benchmark band; diabetes stays well below the naïve aggregate. None of the conclusions tips. Reproduced by `code/common/copas_sensitivity.py`.